\documentclass[14pt]{article}
\pdfoutput=1
\usepackage[table]{xcolor}
\usepackage{cite}
\usepackage{avant}
\usepackage{amssymb}
\usepackage{multirow,booktabs}
\usepackage{caption}
\usepackage{subcaption}
\usepackage{pifont}
\usepackage[mathscr]{eucal}
\usepackage[all]{xy}
\usepackage{listing}
\usepackage{pdfpages}
\usepackage{graphicx} % For including figures
\xyoption{arc}
\xyoption{curve}
\usepackage{bm,amsmath}
\usepackage{braket}
\usepackage{longtable}
\usepackage{mathtools}
\usepackage{hyperref}
\usepackage{breqn}
\usepackage{cite}
\numberwithin{equation}{section}
\usepackage{appendix}
\newcommand\email[4]{#1@#2.#3.#4}
\usepackage{amsfonts}

\usepackage{subcaption}
\usepackage{float}
\numberwithin{equation}{section}
\newcommand\ben{\begin{equation}\nonumber}

		\newfont{\spfnt}{punk12}

		\newcommand\be{\begin{equation}}
		\newcommand\ee{\end{equation}}
		
\begin{document}
	\title{Spatially modulated instabilities of an AdS black hole}
			\author{
				Alisha Gurung\thanks{\email{gurungalisha1508}{gmail}{com}{}}	
				 \, and 
				 Subir Mukhopadhyay\thanks{\email{subirkm}{gmail}{com}{}} 		
			}
			\date{
			 {\small Department of Physics, Sikkim University, 6th Mile, Gangtok 737102.}\\
			 }
			\maketitle
			%\vfil
			\begin{abstract}
				\small
				\noindent
				We analyze instabilities of an Einstein-Maxwell theory obtained from an ${\mathcal N}=2$, D=5 supergravity. The theory admits a gauge Chern-Simons term in presence of which we consider perturbative instability of an AdS black hole solution. We find that the strength of the gauge Chern-Simons coupling saturates the threshold value for which the instability occurs, which we have observed both in the near horizon limit as well as analysis of the normal modes. Inclusion of terms upto fourth order in derivatives admits a mixed gauge-gravitational Chern-Simons term as well. Considering both the Chern-Simons terms, we find that momentum dependent instability sets in below a critical temperature  giving rise to a bell-curve phase diagram which implies this would lead to a spatially modulated solution. We have considered all the terms upto quartic order of the derivatives and discuss the possible approach for further analysis.
				
					\end{abstract}
			\thispagestyle{empty}
			\clearpage

% Table of Contents
\tableofcontents

\section{Introduction}

The AdS/CFT correspondence\cite{Maldacena:1997re,Gubser:1998bc,Witten:1998qj,Aharony:1999ti} has emerged as a powerful framework, offering insights into the nature of strongly correlated systems encountered in condensed matter physics\cite{Hartnoll:2009sz,Zaanen:2015oix}, quantum chromodynamics, quark-gluon plasma, hydrodynamics \cite{Yarom:2009uq,Amoretti:2021lll} and others.  In particular, there are many condensed matter systems, such as high $T_c$ superconductors\cite{Hartnoll:2008kx,Hartnoll:2008vx}, strange metals and insulators \cite{Andrade:2018gqk,Andrade:2017ghg}, whose strongly coupled phenomena are difficult to examine using conventional tools and holographic approach has turned out to be extremely convenient.

From the holographic perspective, if the homogeneous gravitational solutions develop instability that implies a similar instability in the dual theory as well. One of the mechanisms of such instability occur through turning on a large electric field  in a Maxwell-Chern-Simons theory in five dimensions. In those case, so long as the temperature is below a critical, the instability appears, which has a momentum dependence with the value of the momentum within a certain range. As the temperature decreases, the range of the momentum increases leading to a bell shaped curve.  The dependence on the momentum indicates that it will lead to a spatially modulated phase. This kind of spatially modulated instability was studied from the holographic perspective in  \cite{Nakamura:2009tf}, in the context of RN-AdS black hole and the end point of the instability is discussed in \cite{Ooguri:2010kt}. A similar holographic analysis of the instability  for a charged AdS black hole in a $SU(2)\times U(1)$ gauge theory coupled to gravity appeared subsequently\cite{Donos:2011ff}. Its relation to top-down models are also discussed and the endpoint of this instability was numerically obtained in \cite{Donos:2012wi}. We have mentioned about several other studies along similar lines below.  Like the Maxwell-Chern-Simons term in the gauge theory, gravitational theories in five dimensions admit mixed gauge-gravitational Chern-Simons term as well.  A similar study of instability in presence of gravitational Chern-Simons term has appeared for RN-AdS black hole in \cite{Liu:2016hqb}. They also analyse the effects of gauge Chern-Simons term and gauge-gravitational Chern-Simons term.

On the other hand, there are four-dimensional chiral field theory, which exhibits both gauge anomalies and mixed gauge-gravitational anomalies \cite{Alvarez-Gaume:1983ihn}. In the context of field theories, these two kinds of anomalies appeared through the triangle diagram.  If all of the three operators inserted in the diagram have spin one it corresponds to gauge anomaly and the cases where one has spin one while the other two have spin two correspond to mixed gauge-gravitational anomaly.  These anomalies modify the conservation of the current and the energy-momentum and can contribute to the anomaly induced transport \cite{Son:2009tf,Neiman:2010zi,Jensen:2012kj}. There are several effects originating due to mixed gauge-gravitational anomaly that arise in many-body physics at finite temperature \cite{Liu:2016hqb}, such as chiral vortical effect \cite{Landsteiner:2011cp}, odd viscosity \cite{Landsteiner:2016stv} and negative thermal magnetoresistivity \cite{Lucas:2016omy}.
In the holographic context, the mixed gravitational anomaly is encoded in gravity through a gravitational Chern-Simons term \cite{Landsteiner:2011iq}. Therefore, the study of gauge anomaly and mixed gauge-gravitational anomaly from a holographic perspective may illuminate several aspects of instabilities in the dual theories.

Our objective is to explore the possible instabilities in a top-down approach for a gravitational system obtained in the context of string theory/supergravity theory. We will consider a one-charge version of  ${\mathcal N}=2$, D=5 supergravity, which can be obtained from $S^5$ reduction of Type IIB supergravity. The one-charge version admits both the gauge Chern-Simons term and the mixed gauge-gravitational Chern-Simons term, with specific coefficients.  Considering an AdS black hole solution, we explore the instabilities both for the near-horizon geometry as well as for the full black hole solution by examining the normal modes. Since mixed gauge-gravitational Chern-Simons term involves terms of higher-order derivatives, we consider the solution up to the quartic order and obtained linearised equations for the necessary perturbations. However, since it involves derivatives up to fourth order it requires an analysis through canonical formulation, which we briefly mentioned. 

Before we conclude this introduction and move to the next section, let us give a partial brief account of the various relevant studies related to the instabilities leading to spatially inhomogeneous solution. 
In bottom-up approach \cite{Donos:2013wia} considered an Einstein-Maxwell-pseudoscalar model and numerically obtained a black hole solution with spatially modulated horizon with backreaction.  Electrically charged AdS black hole with a neutral pseudoscalar was studied in four dimensions  \cite{Donos:2011bh}.
Spatially modulated instabilities are also found in magnetically charged black branes with AdS$_2$ and AdS$_3$ geometry \cite{Donos:2016hsd}. In a slightly different geometry interpolating between AdS$_2$ and AdS$_4$ having anisotropic Lifshitz scaling and hyperscaling violation in the intermediate with an electric field, a linearised analysis of the instabilities predict a spatially modulated phase as the endpoint \cite{Cremonini:2013epa}. This was further extended to two $U(1)$ gauge fields and found to admit a pair density wave, which involves the intertwined order of  a superconducting phase and a charge density wave (CDW) \cite{Cremonini:2016rbd}. Spectral density, band gap and Fermi surface were explored analysing the Dirac equation in striped phase \cite{Cremonini:2018xgj}. Spatially inhomogeneous configuration as thermodynamically preferred phase has also been shown in \cite{Withers:2013loa}. Formation of inhomogeneous stripes in the bulk and on the boundary was considered in an axionic system \cite{Rozali:2013ama}. Configuration that spontaneously breaks translation symmetry in two directions was considered in checkerboard solution \cite{Withers:2014sja}. Spontaneously broken translation symmetry in a higher derivative model leading to a CDW at T=0 is considered in \cite{Amoretti:2017frz,Amoretti:2017axe}.

A series of works about spatially inhomogeneous models exist in a top-down approach. Spatially inhomogeneous instabilities leading to charge density and spin density waves above a critical density are found in a D7-brane probe in the background of D3-branes with a black hole embedding \cite{Jokela:2014dba}. Similar system of D8-brane probe in the background of the D2-branes were analysed for conductivity and phase structure \cite{Jokela:2011eb}, while an examination of the fluctuations turns out to lead to spatially inhomogeneous striped phase \cite{Jokela:2012se}. At high enough charge density, homogeneous models of D3-D7 branes develop instabilities leading to a combined CDW and spin density wave (SDW) \cite{Jokela:2014dba}. The phase space of the parameters and the nature of the transitions were also studied. Analysis of AC and DC conductivities of this model appears in \cite{Jokela:2016xuy} and it was found by applying an electric field the stripe slides. It was shown in\cite{Jokela:2017ltu} that spontaneously broken spatially modulated ground state with explicit symmetry breaking in the form of ionic lattice gives rise to pinning. Numerical solutions exhibiting spatially modulated phase corresponding to CDW and SDW  was obtained in \cite{Rai:2019qxf} and its Fermi surface has been analysed in \cite{Mukhopadhyay:2020tky}.

The plan of this article is as follows. In Section \ref{sec:1} we briefly review how one can obtain the one-charge version of the gravity theory and the black hole solution. Section \ref{sec:2} consists of exploring instability in the near horizon geometry. In Section \ref{sec:3} we consider the full black hole solution and study the normal modes. Section \ref{sec:4} consists of the higher-derivative action and the equations of linearised perturbations. In Section \ref{sec:5} we briefly discuss the dual field theory and we conclude in Section \ref{sec:6}. An appendix \ref{sec:7} has been added where we briefly review the canonical analysis of a related higher derivative gravity theory.

\section{Single charge black hole solution}
\label{sec:1}
From the perspective of a top-down approach we look for suitable Anti-de Sitter (AdS) black hole solutions, which arise from string theory and/or supergravity theories. A large class of AdS black holes has been discussed in gauged supergravity theories, in the context of AdS/CFT. These gauged supergravities can be obtained as the massless modes of the various Kaluza-Klein compactifications of D=11 and D=10 supergravities. In particular, there are ${\mathcal N}=8$, $D=4$ $SO(8)$ gauged supergravity arise from $D=11$ supergravity on $S^7$, ${\mathcal N}=8$, $D=5$ $SO(6)$ gauged supergravity arise from $D=10$ type IIB supergravity on $S^5$ and ${\mathcal N}=4$, $D=7$ $SO(5)$ gauged supergravity arise from $D=11$ supergravity on $S^4$. These three AdS compactifications, in the absence of black holes, arise from near-horizon geometry of extremal non-rotating $M2$, $D3$ and $M5$ branes. The respective black holes can be interpreted in terms of rotating $M2$, $D3$ and $M5$ branes. 

Once we know the massless sector of the Kaluza-Klein ansatz for compactification of the higher-dimensional theories to the lower-dimensional theories, it is straightforward to embed the lower-dimensional solution to the higher dimension. However, it is quite involved to obtain the correct massless ansatz. For $S^7$, it took many years to finalize while for $S^5$ and $S^4$, the complete ansatz is yet unknown. For the present purpose, one can consider truncations of the gauged supergravity to include only gauge fields in the Cartan subalgebra of the full gauge group, $U(1)^4$, $U(1)^3$, $U(1)^2$ for $S^7$, $S^5$ and $S^4$ compactifications respectively admitting 4-charge AdS$_4$, 3-charge AdS$_5$, 2-charge AdS$_7$ black hole solutions.

In the present article, we will consider the simplest of these three, the $S^5$ reduction of Type IIB supergravity giving rise to ${\mathcal N}=8$, $D=5$ maximal gauged supergravity. The complete details of reduction are not available, but there exist a consistent truncation to ${\mathcal N}=2$ $D=5$  supergravity, with a gauge group given by $U(1) \times U(1) \times U(1)$.
The bosonic sector consists of a graviton, two scalar fields and three gauge bosons ensuing from the $U(1) \times U(1) \times U(1) $ gauge group. 

The complete non-linear Kaluza-Klein ansatz for the compactification of $D=10$, Type IIB supergravity on $S^5$ truncated to $U(1)^3$ is obtained in \cite{Cvetic:1999xp}. As explained in \cite{Cvetic:1999xp}, substituting the reduction ansatz into the equations of motion for the Type IIB theory, we obtain the five-dimensional equations of motion that can be obtained from the following Lagrangian
	\begin{equation} \label{3-charge-action}
	\mathcal{L}= e \hspace{0.05cm} \left( R + 4 g^2 \sum\limits_{i=1}^3 X_i^{-1}  - \frac{1}{4} \sum\limits_{i=1}^3 X_i^{-2}  {F^{(i)}_{\mu \nu}}^2 - \frac{1}{4}\epsilon^{\mu\nu\rho\sigma\lambda} F_{\mu\nu}^1 F_{\rho\sigma}^2 A^3_\lambda  \right)  
	%+  \lambda A_\mu {R^{\beta }}_{\delta \nu \rho} { R^{\delta}}_{\beta \tau \sigma}) \Bigg) 
	\end{equation}	
	$R$ represents the Ricci scalar, the triplet $X_i$ is function of the two scalars, $F^{(i)}_{\mu \nu}$ denotes the field strength tensor for three $U(1)$ gauge fields and $g$ is the inverse of the compact five-sphere which also denotes the gauge coupling constant. The ten-dimensional Bianchi identity $dF_{(5)}=0$ leads to the equations of motion for the scalars and the gauge fields in five dimensions.
	
The above five dimensional Lagrangian admits a three-charge black hole solution whose metric, the scalar triplet \(X_i\) and the gauge field 
\( A^i \) are given by
\be\begin{split}\label{3charge-soln}
ds^2 &= -(H_1 H_2 H_3)^{-2/3} f dt^2 + (H_1 H_2 H_3)^{1/3} (f^{-1}dr^2 + r^2 d\Omega_3^2),\\
 X_i &= H_i^{-1} (H_1 H_2 H_3)^{1/3},\quad A^i = \sqrt{k} (1-H_i^{-1}) \coth{\beta_i} dt\\
 \text{where}\quad f &= k - \frac{\mu}{r^2} + {g^2}{r^2}  (H_1 H_2 H_3), \quad H_i=1+\frac{\mu\sinh^2{\beta_i}} {kr^2},
  \end{split}
\ee
The parameter \( k \) determines the curvature of the transverse space and the value of $k$ can be 1, 0 and -1, where \( k = -1 \) corresponds to the hyperbolic space \( H^3 \), \( k = 0 \) to the flat torus \( T^3 \), and \( k = 1 \) to the positively curved sphere \( S^3 \). For the purpose of our analysis, we focus on the case with \( k = 0 \), where we  absorb the \( k \)-dependence via the rescaling \( k \sinh^2{\beta} \rightarrow  \sinh^2{\beta} \) to obtain
\be
 A^i = \frac{1-H_i^{-1}} {\sinh^2\beta_i} dt, \quad H_i=1+\frac{\mu\sinh^2{\beta_i}} {r^2} .
 \ee

This five dimensional  three charge AdS black hole of \( {\mathcal N}=2 \)  gauged supergravity for k=0 can be embedded in ten-dimension as the decoupling limit of the rotating D3-brane. 
%The non-rotating extremal D3-brane has a decoupling limit in which the space-time of the D3-brane reduces to the product space consisting the five dimensional Anti de Sitter space and a five-sphere, the rotating brane also has a similar decoupling limit given by the AdS black hole, which are given by (\ref{3charge-metric}). At the decoupling limit, the angular momentum associated with the rotating D3-brane are given by $\ell_i^2 = \mu \sinh^2{\beta_i}$. 
The metric and the self-dual five-form field strength supporting the rotating D3-brane are given by
\be \begin{split}
ds_{10}^2 = H^{-1/2} \left( - (1-\frac{2m}{r^4 \Delta})dt^2 + \sum\limits_{i=1}^3 dx_i^2 \right) & +  H^{1/2} \left[ \frac{\Delta dr^2}{H_1H_2H_3 - 2 m r^{-4}} + r^2 \sum\limits_{i=1}^3 H_i ( d\mu_i^2 + \mu_i^2 d\phi_i^2) \right. \\
&\left. - \frac{4 m \cosh \alpha }{r^4 H \Delta} dt \left( \sum\limits_{i=1}^3 \ell_i \mu_i^2 d\phi_i \right) + \frac{2m}{r^4 H \Delta} \left( \sum\limits_{i=1}^3 \ell_i \mu_i^2 d\phi_i \right) ^2\right],\\
\text{where}  \quad \Delta = H_1 H_2 H_3  \sum\limits_{i=1}^3 \frac{\mu_i^2}{H_i} , \quad 
& F_5  = G_5 + ^*\!\!G_5 , \quad  G_5 = dB_4, \\
\text{and} \quad
B_4 = - g^4 r^4 \triangle dt \wedge d^3x &+ \frac{1}{\sinh\alpha} \left( \sum\limits_{i=1}^3  \ell_i^2 \mu_i^2 d\phi_i\right) \wedge d^3x 
\end{split}
\ee

The bosonic Lagrangian given in (\ref{3-charge-action}) can be further truncated down to smaller sector by setting  $X_i=1$ and $F^i_{\mu\nu} = \frac{1}{\sqrt{3}} F_{\mu\nu}$. \cite{Cvetic:1999xp}  This reduces the Lagrangian to
	\begin{equation} \label{1charge-action}
	\mathcal{L}= e \hspace{0.05cm} \left( R + 12 g^2  - \frac{1}{4}  {F_{\mu \nu}}^2 + \frac{1}{12\sqrt{3}}\epsilon^{\mu\nu\rho\sigma\lambda} F_{\mu\nu}^1 F_{\rho\sigma}^2 A^3_\lambda .  \right).  
	\end{equation}	
This action can be embedded in ten-dimensional theories, which was discussed in \cite{Chamblin:1999tk}. In the present note we will consider this simplified one-charge version of this model. 

Chern-Simons terms are abound in supergravity theories and as we observe here, the present one is no exception. In three dimensions Chern-Simons term makes the Maxwell theory massive  and in five dimensions, it becomes tachyonic if an electric field is turned on. As already mentioned in the introduction, due to the electric field, the gauge field mixes with the metric at quadratic order and once the Chern-Simons coupling crosses a threshold value, the geometry becomes unstable for some range of momentum $k$. Such an onset of instability is demonstrated in the context of RN-AdS black hole \cite{Nakamura:2009tf,Ooguri:2010kt} below a critical temperature . These instabilities occur for a range of momenta that typically excludes $k=0$. In the dual theory, this instability signals a novel phase transition at a finite chemical potential where the charge current develops a position-dependent expectation value.  Such instabilities appear in a variety of models \cite{Donos:2011ff,Donos:2012wi} and usually it leads to a spatially modulated solution as its endpoint. Therefore it is interesting to study the possible instabilities in this one-charge version of this model.

%%%
While we are examining the instabilities triggered by topological terms such as Chern-Simons coupling in the gauge theory, we can enquire about other possible such terms that may occur in the present theory. The natural term to include in this discussion is the mixed gauge-gravitational Chern-Simons term. The instability triggered by the gauge-gravitational Chern-Simons term has been demonstrated in the context of  RN-AdS black hole \cite{Liu:2016hqb}. But the gauge-gravitational Chern-Simons terms are higher order in derivatives and will not appear if we restrict the action up to terms quadratic in derivatives.  For the time being we will consider only the gauge-gravitational Chern-Simons term among the terms which are higher order in derivatives, because this will play a qualitatively non-trivial role in the discussion of stability and because one can argue that the gauge-gravitational Chern-Simons can be treated non-perturbatively unlike the other higher derivative terms.

We begin with the Lagrangian, given in \refeq{1charge-action} along with the mixed gauge-gravitational Chern-Simons term to obtain,
\begin{equation}
	\mathcal{L}= e \hspace{0.05cm} \Bigg( -R+12 g^2 - \frac{1}{4}  {F_{\mu \nu}}^2+ \epsilon^{\mu \nu \rho \tau \sigma} (- \frac{\alpha}{3 }A_\mu F_{ \nu \rho } F_{\tau \sigma }   +  \lambda A_\mu {R^{\beta }}_{\delta \nu \rho} { R^{\delta}}_{\beta \tau \sigma}) \Bigg) ,\label{eq:1}
\end{equation}	
We have introduced a general parameter $\alpha$ to denote the strength of the gauge theory Chern-Simons term, which has a specific value given by $\alpha=  \frac{1}{4 \sqrt{3}}$ \cite{Cvetic:1999xp,Cremonini:2008tw}. $\lambda$ represents coefficient of gravitational Chern-Simons term, and its value is \( \frac{c_2}{24} \left( \frac{1}{16\sqrt{3}}\right)\) \cite{Cremonini:2008tw}.  \( g \) is the coupling constant that characterizes the strength of the interaction. There are scalar fields as well but, since they are not playing any crucial role in the present analysis, they have been omitted. 

The equations of motion following from the above Lagrangian can be organised in Maxwell and Einstein equation which are given by,
\begin{equation}
\begin{split} \label{eom}
\nabla_\nu F^{\nu \mu}  + \epsilon^{\mu \beta \rho \lambda \sigma}(\alpha F_{\beta \rho} F_{ \lambda \sigma}  +  \lambda {R^\delta}_{\eta \beta \rho}  {R^\eta}_{\delta \lambda \sigma} ) & = 0 \\       
-R_{\mu \nu} - \frac{1}{2} g_{\mu \nu} \left(  -R -\frac{1}{4}  F^2 +12 g^2    \right) - \frac{1}{2} F_{\mu \lambda} {F_{\nu}}^\lambda  - 2 \lambda \epsilon_{\tau \eta \lambda \sigma ( \mu | } \nabla_\beta (F^{\eta \tau}{ {R^\beta}_{| \nu)}}^{\lambda \sigma}) &= 0     \end{split}
\end{equation}	

We consider the single charge analog of the three-charge black hole solution as given in \eqref{3charge-soln}, which represents a five-dimensional  asymptotically AdS black hole \cite{Cremonini:2008tw},
The metric  is given by,
\begin{equation}
\begin{split}    \label{1-charge-bh}
%{eq:2},
ds^2 &=  \frac {f}{H^2} dt^2 - H \left( \frac{1}{f}dr^2 + r^2 d x^2+ r^2 d y^2+r^2 d z^2   \right) ,\\
	f&= k- \frac{\mu}{r^2} + g^2 r^2 H^3 , \quad	 H = 1 + \frac{Q}{r^2}.\\
	A &= \sqrt{\frac{3(k Q+ \mu)}{Q}}\left(1-\frac{1}{H} \right)  \hspace{0.05cm}dt. 
	\end{split}
\end{equation}
Here \( f \) denotes the blackening function. \( \mu \) is the chemical potential and \( Q \) characterizes the charge and is given by \( Q =  \mu^2 \sinh^2{\beta}/k \). From a higher-dimensional perspective such charge terms can be interpreted as angular momenta, corresponding to rotating D3-brane configurations\cite{Cvetic:1999xp}. 	
 \( A\) represents the gauge field.
As mentioned earlier, we focus on the simplest case with \( k = 0 \), where we allow to absorb the \( k \)-dependence via the rescaling \(  \mu^2 \sinh^2{\beta} \rightarrow  k \mu^2 \sinh^2{\beta} \).
Substituting the above scaling and k value in equations~\eqref{1-charge-bh}, yield the following simplified forms, 
\begin{align*}
	A= \sqrt { \frac{3 \mu }{Q} } \left(1-\frac{1}{H} \right) dt  \hspace{0.05cm} , & \quad \quad  \quad  f= - \frac{\mu}{r^2} + g^2 r^2 H^3  
\end{align*}

This completes the description of the black hole solution, in the single $U(1)$ charge theory, which can be obtained through KK reduction of higher-dimensional supergravity theory which will be the focus of our interest. In section \ref{sec:2} and we will analyze the near horizon behavior, linearized perturbations and relevant implications from the resulting solutions.

\section{Near horizon analysis} 	
\label{sec:2}

As a first step towards analysis of stability of the black hole solution given in \eqref{1-charge-bh}, we begin with an examination of its near-horizon geometry in the zero-temperature limit. In this limit, the geometry simplifies to \( AdS_2 \times \mathbb{R}^3 \) configuration. In the context of gauge Chern-Simons term in RN-AdS black hole solution it turns out that the near-horizon analysis gives a sufficient but not a necessary condition since there are normalizable modes in the full solution, which does not reduce to the normalizable mode in the near-horizon limit.  The near-horizon analysis along with both the Chern-Simons term for the present black hole turns out to be very similar to the one conducted in \cite{Liu:2016hqb} for RN-AdS black hole except that our analysis contains in addition the coupling constant \( g \) as a variable parameter. This will be useful to capture the essential physics and set the stage for the analysis for normal mode. 

In order to find the near horizon limit, we denote \( r_h \) as the value of the radial coordinate at the horizon and rewrite the metric in terms of $(r - r_h)$. For convenience we introduce instead its inverse as \( \rho = {1}/{(r - r_h)} \), so that \( \rho \to 0 \) corresponds to the asymptotic boundary of the bulk gravitational background while the horizon lies at \( \rho \to \infty \) . At this limit the solution reduces to 
\begin{equation}%eq:6
ds^2 = \frac{L^2}{\rho^2} (d \hat{t}^2- d\rho^2)- d \hat{x}^2-d \hat{y}^2-d \hat{z}^2 , \quad  \quad A_t = -\frac{1}{\sqrt{6} g \rho} , \label{nh-soln} %\label{nh-metric}
 \end{equation}
 where the radius squared of the AdS$_2$,  \( L^2 \)  is given by
 \begin{align*}
	L^2= \frac{1}{12g^2} 
	\end{align*}
 We are interested in the stability of this near-horizon solution by turning on small perturbations around this background and examining the associated mass matrix. In flat spacetime, if the mass squared associated with any mode acquires negative value, that. signals a tachyonic instability. In Anti-de Sitter (AdS) geometry, the effect of the curvature of spacetime allows negative values for mass-squared and the lower bound is pushed down to Breitenlohner-Freedman (BF) bound. As long as the value of the mass squared remains above BF bound, the configuration remains  stable despite being tachyonic in the Minkowski spacetime \cite{Gubser:2001zr}.  The BF bound defines the allowed range of masses for stable modes \cite{Breitenlohner:1982jf} and  we need to examine whether, the BF bound is satisfied by the eigenvalues of the mass matrix. In the present case the BF bound turns out to be, \( M^2 = -3 g^2 \)\cite{zaanen_liu_sun_schalm_2015}. In order to procure instability in the near-horizon, this bound must be violated, i.e., some eigenvalue of the mass matrix should satisfy 
 \be M^2 \leq -3 g^2 .\label{BF} \ee

 Our general approach will be to analyse whether the present near horizon solution is susceptible to instability due to any small fluctuation. In order to examine it we consider small perturbations to both the metric and gauge field\cite{Liu:2016hqb} as follows:
\begin{equation} \label{perturbation}
 \delta g_{ai} = h_{ai} (t, \rho )e^{ikx}  %eq:9     
 \quad,\quad
 \delta A_{i} = a_{ i} (t, \rho )e^{ikx}       %eq:10}
 \end{equation}
 
Here, \(a\) and \(i\) correspond to the coordinates \((t, x)\) and \((y, z)\), respectively. By substituting the linearized perturbations~\eqref{perturbation} into the equations of motion~\eqref{eom}, we obtain the following equations. Substituting the perturbations \eqref{perturbation} into the Maxwell's equations, the first of \eqref{eom} leads to
 \begin{equation}
  12 g^2 \rho^2(a_i '' -{\partial_t}^2 a_i)-k^2 a_i -16  \sqrt{6} i g k \alpha \epsilon^{i j} a_j  - 24 \sqrt{6} g^3 \rho^2 (h_{ti}' + 8 \sqrt{6} i g k \lambda  \epsilon^{ij} h_{tj}')=0.   \label{Max-pert}  %\label{eq:11}
  \end{equation}
  Similarly substituting into Einstein's equation, the second of \eqref{eom} leads to
  \begin{equation}
\begin{aligned}
 & -6 g^2 \rho^2 h_{ti}'' -12 g^2 \rho h_{ti}'+\frac{k^2}{2} h_{ti}+\frac{i k}{2} \partial_t h_{xi} + \sqrt{6} g a_i'  +\\
  &\quad+ i 4 \sqrt{6} k \lambda \epsilon^{ij}(12 g^3 \rho^2 h_{tj}''+24 g^3 \rho h_{tj}'- k^2 g  h_{tj}-ik g \partial_t h_{xj}+2\sqrt{6} g^2 a_j') = 0   \label{Ein-pert1}%  \label{eq:12} 
  \end{aligned}
\end{equation}
 \begin{equation}
  h_{xi}'' - {\partial_t}^2 h_{xi} +i k \partial_t h_{ti}-i 8 \sqrt{6}g k \lambda  \epsilon^{ij} (h_{xj}''-{\partial_t}^2 h_{xj}+ik\partial_t h_{tj})=0          \label{Ein-pert2}% \label{eq:13}
  \end{equation}
 \begin{equation}
\sqrt{6} i g \partial_t a_i-6 i g^2 \rho^2 \partial_t {h_{ti}}' -\frac{k}{2} h_{xi}'+i 4 \sqrt{6} k \lambda \epsilon^{ij}(k h_{xj}'+12 g^3 \rho^2 \partial_t h_{tj}'+2 \sqrt{6}  g^2 i \partial_t a_j)=0         \label{Ein-pert3}%  \label{eq:14} 
\end{equation}

In order to check the instability we need to reorganize the equations by performing appropriate field redefinitions following procedure similar to  \cite{Liu:2016hqb}, which allow us to express these equations in a more tractable form. 
The relevant field redefinitions are given by:
 \begin{equation}\begin{split}
\phi_i &= a_i + i 8 \sqrt{6} g k \lambda \epsilon^{ij} a_j - \sqrt{6} g \rho^2 (h_{ti}' -i 8 \sqrt{6} g k \lambda \epsilon^{ij}h_{tj}')  \label{field-redef} \\ %{eq:15} 
\Phi_b &=(a_y+i a_z, a_y- i a_z, \phi_y+i \phi_z, \phi_y- i \phi_z)  %\label{eq:16} 
\end{split}\end{equation}
  
 As a result of redefinition, we obtain a simpler set of coupled second-order differential equations given below,
\begin{equation}
\begin{split}
 &12 g^2 \rho^2 (\Phi_{(1,2)}'' - {\partial_t}^2\Phi_{(1,2)}) +\left( -k^2+ 72 g^2 \mp 16 \sqrt{6}g k\alpha \pm 192 \sqrt{6} g^3 k \lambda - \frac{96 g^2}{1\mp 8 \sqrt{6} g k \lambda }\right) \Phi_{(1,2)} + \\
  &\quad - 24 g^2 \left(1-\frac{2}{1\mp 8 \sqrt{6} g k \lambda }  \right)  \Phi_{(3,4)}=0       \label{field-redef1} \\ %{eq:17}
&12 g^2 \rho^2 (\Phi_{(3,4)}'' - {\partial_t}^2\Phi_{(3,4)})  + (k^2 \pm 8 \sqrt{6} g k^3 \lambda) \Phi_{(1,2)}-k^2 \Phi_{(3,4)} =0   %\label{eq:18} 
\end{split}\end{equation}

We now proceed to write the coupled equations  in a form analogous to a mass matrix \( M^2 \). For this purpose, we set \( \Phi_+ = \left( \begin{smallmatrix} \Phi_1 \\ \Phi_3 \end{smallmatrix} \right) \) and \( \Phi_- = \left( \begin{smallmatrix} \Phi_2 \\ \Phi_4 \end{smallmatrix} \right) \), which allows us to write the equations in a matrix form as follows,

\begin{equation} 
(12 g^2 \rho^2(\partial_\rho^2-\partial_t^2)- M_\pm^2)\Phi_\pm = 0       \label{eq:19}
\end{equation}

where

\[ M^2_\pm =
\begin{bmatrix}
k^2- 72 g^2 \pm 16 \sqrt{6}g k\alpha \mp 192 \sqrt{6} g^3 k \lambda + \frac{96 g^2}{1\mp 8 \sqrt{6} g k \lambda } &  24 g^2 \left(1-\frac{2}{1\mp 8 \sqrt{6} g k \lambda }  \right) \\
 -(k^2 \pm 8 \sqrt{6} g k^3 \lambda) & k^2
\end{bmatrix}
\]
 The effective mass-squared of the perturbations is extracted from the associated mass matrix derived in the AdS background. In our setup, the BF bound is given in \eqref{BF} 
  Considering the coefficients of the Chern-Simons term we can have three different cases, which we briefly mention in the following.

  \begin{itemize}
 \leftskip=1cm
  
 \item {\bf Gauge Chern-Simons term:} When only the gauge Chern-Simons term (with coefficient \( \alpha \)) is turned on, we find there is a critical value of the coefficient exists, beyond which the BF bound is not violated and the instability sets in. It turns out that this critical value is approximately \( \alpha_c = 1/4 \sqrt{3} \), which matches exactly with the coefficients of the Chern-Simons term in our action obtained in the top-down approach. In other words, our solution is marginally stable. It may be interesting to understand whether this precise agreement has any relation with the supersymmetry.  

\item {\bf Gauge-gravitational Chern-Simons term:} On the other hand, turning on only the gravitational Chern-Simons term alone does not lead to a violation of the BF bound and so it does not generate an instability on its own.

\item {\bf Both Gauge and gauge-gravitational Chern-Simons terms:} In contrast, when both gauge and gravitational Chern-Simons terms are simultaneously turned on, even with a small coefficient for the gravitational term, the BF bound is violated (e.g., for \( g = 1 \), as used in certain conventions). This signals the appearance of an unstable mode in the bulk. 

\end{itemize}
\leftskip=0cm

The system thus undergoes a transition from a stable to an unstable phase depending on the interplay between the gauge and gravitational Chern-Simons terms. Moreover, as the coefficient of the gravitational Chern-Simons term increases, the violation of the BF bound becomes more significant, indicating stronger instability within the bulk geometry.

\section{ Normal modes %with inclusion of Gravitational Chern-Simons term 
} 
\label{sec:3}

The observation of BF bound violation in the near-horizon regime of the previous Section \ref{sec:2}, by the inclusion of gravitational Chern-Simons term, suggests the possible occurrence of instability in the full black hole geometry. 
As  one can observe in the context of near horizon geometry that the metric fluctuations of type $h_{xi}$ are related to time derivative of other fluctuation. Therefore restricting ourselves to the static cases, we can consider the metric fluctuations of type $h_{ti}$. Thus, we extend our analysis by introducing appropriate perturbations 
\cite{Liu:2016hqb}, \begin{equation} 
\delta (ds^2)= 2 Q(r)( cos(kx) dt dy+ sin(kx) dt dz ), 
\quad \quad  \quad \quad  \delta A= a(r) ( cos(kx) dy+ sin(kx) dz ),
\label{eq:20} \end{equation}
where $Q(r)$ and $a(r)$ are the perturbations in the metric and the gauge field respectively. We substitute the perturbation ansatz \eqref{eq:20} into the general equations of motion \eqref{eom}, corresponding to the full black hole background and the gauge field and obtain linearised equations for $Q(r)$ and $a(r)$.

By substituting in the Maxwell's equation, the first of \eqref{eom}, we obtain
\begin{equation}
\begin{split}
& f H^3 r^4 a''+ 
H^2 r^3\Big[ H r f' + f \left(H - r H'\right)\Big) a'  - H^3 k r^2 \left(k + 8 H r \alpha A'  \right] a + \\
& - 2 \Big[
  H^5 r^4 A' 
  +2 H k r \lambda \Big\lbrace6 f H r H' 
  + 15 f r^2 (H')^2 
  - H r f' (4 H + 9 r H') + 2 H (2 f H + H r^2 f'' - 3 f r^2 H'')\Big\rbrace \Big] Q'  +\\
& +2  \Big(
  16 f H^3 k \lambda 
 +32 f H^2 k r \lambda H' 
  + 72 f H k r^2 \lambda (H')^2 
  +30 f k r^3 \lambda (H')^3 +\\
&  - H^4 r^3 A' \left(H + r H'\right) 
  - 2 H k r \lambda f' \left(8 H^2 + 22 H r H' + 9 r^2 (H')^2\right) 
  - H^5 r^4 A'' +\\
& + 8 H^3 k r^2 \lambda f'' 
  + 4 H^2 k r^3 \lambda H' f'' 
  - 24 f H^2 k r^2 \lambda H'' 
  - 12 f H k r^3 \lambda H' H'' \Big) Q  =0          \label{eq:21}
\end{split}
\end{equation}

A similar equation follows from the Einstein's equation, the second of \eqref{eom}.
\begin{equation}
\begin{aligned}
& \quad 6 f H^5 r^2 \left(r - 4 k \lambda A'\right) Q'' -6 f H^4 r^2  \Big( -\left(r - 4 k \lambda A'\right) H' + H \left(-1 + 4 k \lambda A''\right) \Big) Q' +\\
& + 2 H^3  \Big(
  H^4 r^3 (A')^2 
  + 3 r \Big(
    -2 f r^2 (H')^2 
    + H r f' (2 H + r H') 
    + 4 f H r H' (-1 + k \lambda A'')+ \\
& - H^2 \left(k^2 + 8 g^2 H r^2 - 8 f k \lambda A'' \right)
  \Big) 
  + 12 k \lambda A' \left(
    6 f H r H' 
    + f r^2 (H')^2 
    + H \left(H (2 f + k^2) + f r^2 H'' \right)
  \right)
\Big) Q + \\
& - 3 f H  \Big(
  H^4 r^3 A' 
  + 2 k \lambda \left(
    6 f H r H' 
    + 15 f r^2 (H')^2 
    - H r f' (4 H + 9 r H') 
    + 2 H (2 f H + H r^2 f'' - 3 f r^2 H'')
  \right)
\Big) a' + \\
& + 6 f k \lambda \Big(
  -24 H r (f + r f') (H')^2 
  + 30 f r^2 (H')^3 
  + H^2 r \left(-2 f''' H r + 6 f H''' r + 3 (2 f + 5 r f') H'' \right)+\\
& + 3 H H' \left(4 H r f' + 3 H r^2 f'' - 2 f (H + 6 r^2 H'') \right)
\Big)  a  =0    \label{eq:22}
\end{aligned}
\end{equation}

We will use these linearized equations of motion to analyze the behavior of normal modes whose existence may signal spatially modulated  instabilities. To extract these modes, we need to solve the coupled second-order differential equations \eqref{eq:21}\eqref{eq:22}, with appropriate boundary conditions both at the asymptotic boundary and horizon of the black hole.
 
 We begin with the boundary condition to be imposed at the horizon. We consider a near-horizon expansion of the fields, in terms of \(\sigma\) at the limit \( r \rightarrow (r_h + \sigma) \), where \( \sigma \) is a small parameter, It is sufficient to consider only a few terms for our purpose and we set
\begin{equation}
Q = q_0 +  q_1 \sigma +q_2 \sigma^2+.... ,       \quad\quad \quad \quad \quad  a = a_0 +  a_1 \sigma +......         .  \label{eq:23}
\end{equation}
Plugging in these expansions in \eqref{eq:21}\eqref{eq:22} and trying to solve the equations in orders of \(\sigma\) leads to the following result.
 \begin{equation}
 \begin{split} \label{eq:24}%\label{eq:25}
 q_0 &= 0,\\   
  q_2 & = 
\frac{1}{8 \pi T} \Bigg( \frac{q_1}{(r_h^3 + r_h \mu^2 \sinh^2 \beta)^2} \Bigg[ 
														r_h^4 (k^2 + 4 r_h (2 g^2 r_h - 3 \pi T))\\
														&+ 2 \mu^2 (r_h^2 (k^2 + 4 r_h (3 g^2 r_h - \pi T)) + 4 \mu) \sinh^2 \beta 
														+ (k^2 + 4 r_h (6 g^2 r_h + \pi T)) \mu^4 \sinh^4 \beta+ 8 g^2 \mu^6 \sinh^6 \beta 
															\Bigg]\\
&- \frac{a_0 k \mu^{3/2} \sinh \beta(-48 \alpha \mu^{3/2} \sinh \beta + \sqrt{3} k (r_h^2 + \mu^2 \sinh^2 \beta))}{r_h (r_h^2 + \mu^2 \sinh^2 \beta)^3} \Bigg)  ,   \\     
a_1 &= - \frac{\sqrt{3} \, q_1 \, \mu^{3/2} \, \sinh\beta}{\pi \, r_h^3 \, T}
+ \frac{a_0 \, k \left( k \, r_h^2 - 16 \sqrt{3} \, \alpha \, \mu^{3/2} \, \sinh\beta + k \, \mu^2 \, \sinh^2\beta \right)}{4 \pi \, r_h^2 \, T \left( r_h^2 + \mu^2 \, \sinh^2\beta \right)}  
\end{split}
\end{equation}
So the parameter $q_0$ turns out to be zero and all other parameters are dependent on \(a_0\) and \( q_1\).
 
 On the other hand, in the  asymptotic  region \( r \to \infty \),  we consider the behavior of the fields near the boundary of the bulk geometry. Closely examining the equation, we see the asymptotic behaviour are given by,
\begin{equation}
Q= c_0 r^2 +\frac{c_1}{r^2}+\frac{c_2}{r^4}+\frac{c_3}{r^6}+..........      %\label{eq:26}
\quad,\quad\quad\quad\quad
a= b_0 r^2 + \frac{b_1}{r^2}+\frac{b_2}{r^4}+\frac{b_3}{r^6}+.......... 
       \label{eq:27}
\end{equation}
As one can observe, both \( Q \) and \(a\) has asymptotic divergences and in order to have normal modes we make them convergent by imposing the conditions \( b_0 = 0 \) and \( c_0 = 0 \)\cite{Liu:2016hqb}. 

With this, the boundary conditions are now specified both at the asymptotic boundary and at the horizon, in terms of the parameters given by \( (a_0, q_1, b_1, b_2, b_3, c_1, c_2, c_3) \).
Thus it has been reduced to a boundary value problem with the coupled second-order differential equations \eqref{eq:21} and \eqref{eq:22}, with the boundary conditions as specified by the parameters \( (a_0, q_1, b_1, b_2, b_3, c_1, c_2, c_3) \).

As analytic solutions of such boundary value problem is not easy, we employ a numerical method known as the double shooting method\cite{Liu:2016hqb}. This technique involves integrating the differential equations from both ends, from the horizon (initial point) and from the asymptotic boundary (final point), using the respective boundary conditions. If the solutions from both directions align smoothly within non trivial spatial range, it imply that the boundary conditions are satisfied, and the solutions are valid. One method to examine this alignment of two sets of solution by considering the Wronskian\cite{Krikun:2013iha,Kiritsis:2015hoa}, a determinant constructed from the solutions and their derivatives. If the Wronskian vanishes, it confirms that the solutions are consistent and the boundary conditions are met.

To identify the normal modes of the system, we have adopted this method of examining the Wronskian. In more concrete terms, we analyze the Wronskian constructed from two independent solutions of the linearized equations \eqref{eq:21} and \eqref{eq:22}, vanishes at a some specific points. This vanishing indicates that the solutions from the horizon and the boundary connect smoothly, satisfying both boundary conditions, and thus correspond to normal modes. We have analysed the existence of the normal mode for a wide range of values of spatial momentum $k$ and the temperature $T$. At a given temperature, we find that there is only a specific range of \(k\) for which the Wronskian vanishes, signalling presence of the normal modes. As the temperature $T$ increases, the range of $k$ values for which the normal mode exists diminishes, eventually reaching a critical temperature beyond which no normal modes are found. This behavior results in a bell-shaped curve in the $T$-$k$ plane, which is the characteristic behavior found in various other contexts of instabilities leading to spatial modulation.
We perform this analysis for two distinct cases: first, considering only the gauge Chern-Simons term, and second, incorporating both the gauge and gravitational Chern-Simons terms:

\begin{itemize}

\item {\bf Gauge Chern-Simons term only:}

To begin with we set \(\lambda\) equal to zero to turn off the gauge-gravitational Chern-Simons term and consider only the gauge Chern-Simons term. As we have already observed in other contexts including the near- horizon limit, it turns out that if the coefficient of the Chern-Simons term $\alpha$ is too small, it does not admit any normal mode, despite varying other parameters such as the chemical potential \(\mu\), the coupling constant \(g\), and other parameters derived from the boundary conditions. There is a critical value of $\alpha$ beyond which the normal modes will begin to appear.  Once again this critical values is \(\alpha_c = {1}/{4\sqrt{3}}\) and as we mentioned earlier it is interesting to see that this is exactly the value of \(\alpha\) of our single charge model, derived from the \({\mathcal N} =8\), \(D=5\) gauged supergravity, a fact which  could be related to the supersymmetry.

\begin{figure}[H]
    \centering
    % First image
    \begin{subfigure}[b]{0.49\textwidth}
        \includegraphics[width=\linewidth]{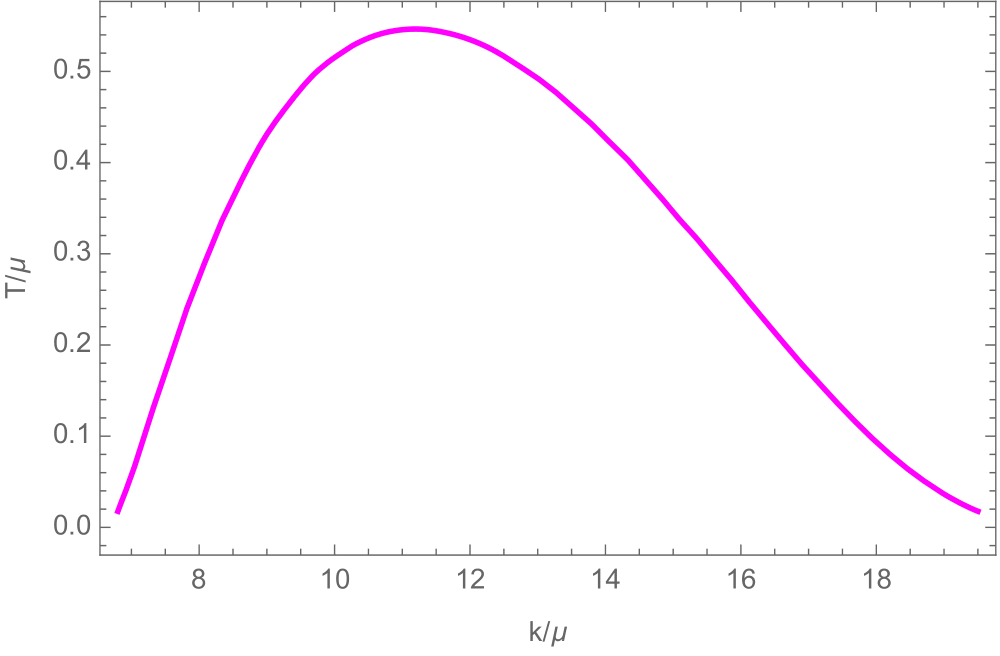}
        \caption{$\mu=0.1$, g = 1}
        \label{fig1a}
    \end{subfigure}
    \hfill
    % Second image
    \begin{subfigure}[b]{0.49\textwidth}
        \includegraphics[width=\linewidth]{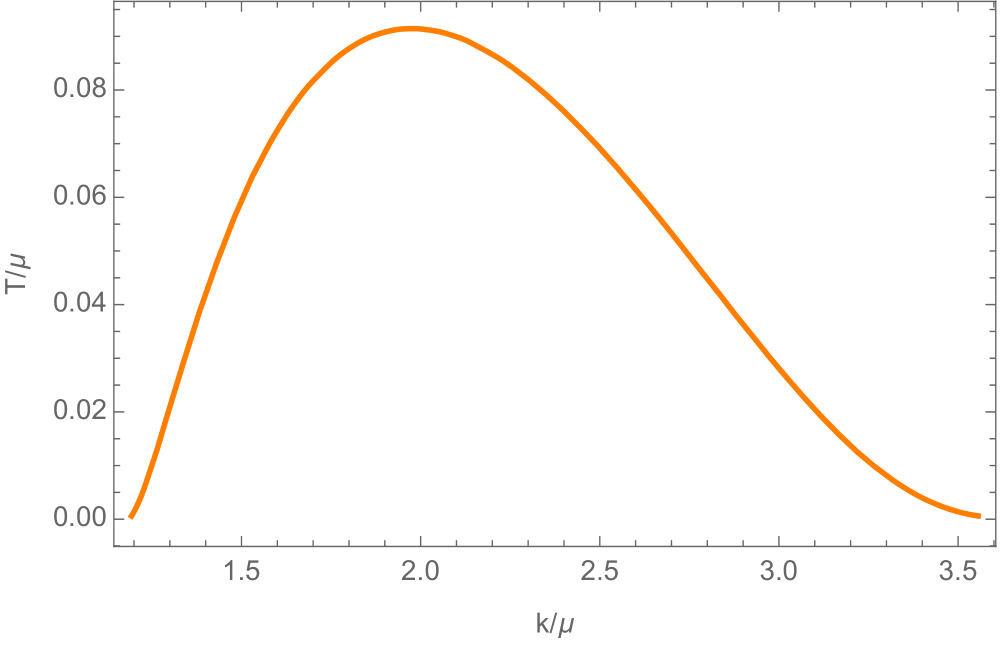}
        \caption{$\mu=1$, g = 0.98}
        \label{fig1b}
    \end{subfigure}
      \vspace{.5in}
      
          % First image
    \begin{subfigure}[b]{0.49\textwidth}
        \renewcommand\thesubfigure{c} % Change label to (c)
        \includegraphics[width=\linewidth]{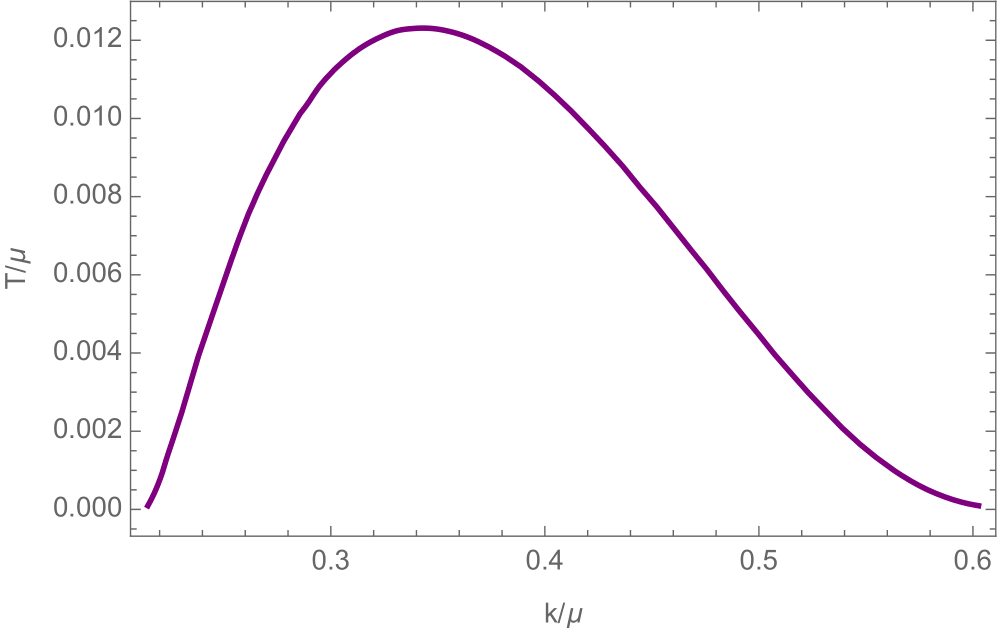}
        \caption{$\mu=10$, g = 0.9}
        \label{fig1c}
    \end{subfigure}
    \hfill
    % Second image
    \begin{subfigure}[b]{0.49\textwidth}
        \renewcommand\thesubfigure{d} % Change label to (d)
        \includegraphics[width=\linewidth]{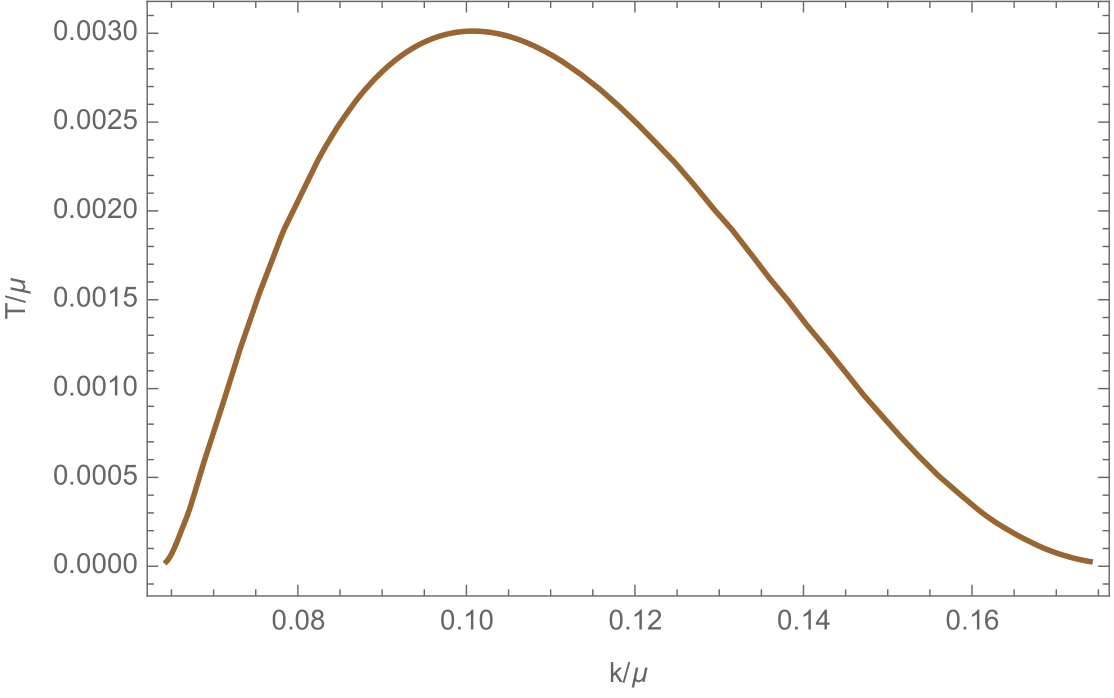}
        \caption{$\mu=50$, g = 0.85}
        \label{fig1d}
    \end{subfigure}
    \caption{Bell curves for four different sets of $\mu$ and $g$, with fixed $\lambda = 0.01$.}
    \label{fig2}
\end{figure}

\item {\bf Both Gauge and Gravitational Chern-Simons term: }

Next we incorporate both the gravitational and gauge Chern-Simons terms and obtain the range of momentum \(k\) as a function of the temperature \(T\).
We consider the characteristic bell-shaped curve in the \(T\)-\(k\) plane for different values of chemical potentials and coupling constants. In more concrete terms we set  \(\lambda=.01\), and analyse the bell curves for the following sets:
\ben ( \mu = 0.1, g =1),\quad ( \mu = 1, g = 0.98),\quad  (\mu = 10, g = 0.9), \quad(\mu = 50, g = 0.85). \ee 
The results are assembled in  Fig. \ref{fig1a}, \ref{fig1b}, \ref{fig1c}, \ref{fig1d}.

We have normalised both \(k\) and \(T\) with the chemical potential \(\mu\). These extend our investigation by examining how this bell-shaped curve evolves with different values of \(\mu\). As one can observe as the chemical potential increases, the normalised range of momentum and temperature decreases monotonically.

Extending our analysis, we investigate the influence of varying the gravitational Chern-Simons coupling coefficient $\lambda$, on the normal modes. By fixing the chemical potential \(\mu\) and the coupling constant \(g\), we compute the bell-shaped curves in the \(T\)-\(k\) plane for three distinct values of \(\lambda\): \(0.04\), \(0.01\), and \(0.005\), which are given in Fig.\ref{fig:lambda0006}. Our findings reveal that as \(\lambda\) decreases, the extent of the unstable region in the \(T\)-\(k\) plane contracts. This behavior prove that the gravitational Chern-Simons term in driving the instability as larger values of $\lambda$ enhance the instability, while smaller values suppress it.
\end{itemize}

\begin{figure}[H]
    \centering
    %Fourth image
    \includegraphics[width=\linewidth]{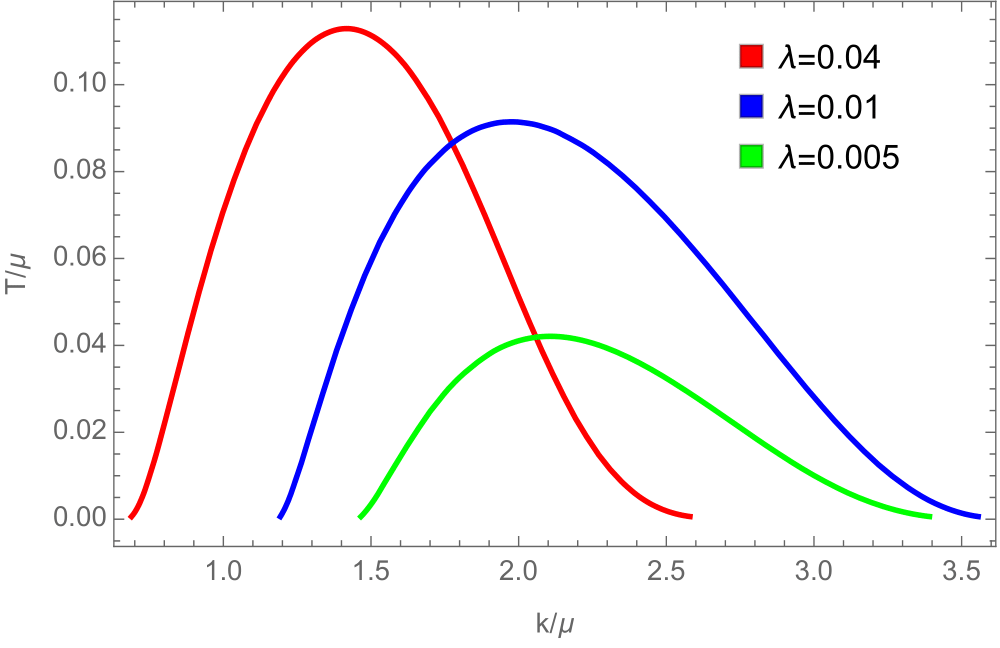}
    \caption{Bell curves for $\mu = 1$, $g = 0.98$, and different values of $\lambda$: $0.04$, $0.01$, and $0.005$}
    \label{fig:lambda0006}
\end{figure}

To summerise, we consider various combinations of the chemical potential \(\mu\), the coupling constant \(g\) and the gauge-gravitational Chern-Simons coefficient \(\lambda\). We observe the emergence of a well-defined bell-shaped curves in the \(T\)-\(k\) plane depicting the temperature-dependent range of momenta \(k\) where the Wronskian vanishes, indicating the presence of normal modes within this region and thus confirming the existence of spatially modulated unstable modes in the bulk gravity. 
%We find \(\lambda\)  plays a significant role  in inducing instabilities within the gravitational bulk as variations in \(\lambda\) lead to substantial changes in the shape and extent of the bell curve.  When both the gravitational and gauge Chern-Simons terms are included, instabilities become more prominent. 

\section{Higher derivative corrections} 
\label{sec:4}

In the earlier section we have studied the stability of the black hole solution as well as its near horizon version in presence of both gauge and gauge-gravitational Chern-Simons terms. The gauge-gravitational Chern-Simons term is a four-derivative term but we restrict our action upto quadratic terms only in the spirit that the the gauge-gravitational Chern-Simons term  is non-perturbative and unlike the other higher derivative terms and it plays a crucial role in the analysis of stability.  Consistency requires inclusion of other four-derivative terms, which may lead to significant modifications in the system’s behavior and its underlying dynamics. In view of that in this section we consider the action upto quartic terms.

There are ample reasons to consider higher-derivative terms in a gravity theory.  Gravity theories  as well as supergravity theories are considered to work successfully as an effective field theory, which is a low energy limit of a UV complete theory, e.g. string theory. Therefore, a natural extension is to consider higher derivative corrections as string theory leads to high order derivative correction, e.g. $R^2$ and $R^4$ correction, that occur in heterotic and Type II supergravity respectively and it has been argued that the absence of $R^2$ term is due to maximal supersymmetry. So, it is legitimate to consider stringy corrections to two-derivative theory, which are higher order in derivatives. 
Another rationale to include such corrections follows from the perspective of AdS/CFT, as this may teach us more about finite couplings as well as 1/N effects in the dual  \( {\mathcal N}=1 \) super-Yang-Mills theory. 
We will briefly review the construction of the four-derivative action following \cite{Cremonini:2008tw}, which may be consulted for details.

Since the present model corresponds to five dimensional  \( {\mathcal N}=2 \)  gauged supergravity we can expect the first correction occurs at $R^2$ order and restrict ourselves to four-derivative terms only. 
As argued in \cite{Cremonini:2008tw} one can expect to have four-derivative terms uniquely fixed by supersymmetry. The ungauged theory may follow from M-theory compactified on a Calabi-Yau three-fold with higher derivative correction
\begin{equation}
e^{-1} \delta\mathcal{L} = \frac{c_{2I}}{24} \left( \frac{1}{16}  \epsilon_{\mu\nu\rho\lambda\sigma} A^\mu R^{\nu\rho\alpha\beta } R^{\lambda\sigma}_{\hphantom{\lambda\sigma}\alpha\beta}  + ... \right) ,\label{eq:0}
\end{equation}	
Comparing this term with the Calabi-Yau reduction of the M5 brane anomaly term shows that \(c_{2I}\) represents second Chern class on the Calabi-Yau manifold. In our case we are interested in the gauged supergravity case which corresponds to Type IIB string theory compactified on AdS\(_5\times\text{Y}^5\), where \(\text{Y}^5\) is Sasaki-Einstein, which is dual to \( {\mathcal N}=1 \) super-Yang-Mills theory in 4 dimensions. In this case their stringy origin is less clear but the \(c_{2I}\) is related to the gauge theory data using holographic anomaly. 

The conventional on-shell formulation of minimal ${\mathcal N}=2$ gauged supergravity is given in terms of the graviton multiplet consisting of the metric $g_{\mu\nu}$, a graviphoton $A_\mu$ and a symplectic-Majorana spinor $\psi_\mu^i
$ with $i=1,2$ labelling the doublet of $SU(2)$. The two-derivative Lagrangian consisting of the bosonic degrees of freedom is given by \eqref{1charge-action}. In the present case we are interested in the four derivative correction of the Lagrangian consistent with supersymmetry. There are  purely gravitational terms e.g. \(R^2, R_{\mu\nu}R^{\mu\nu}, R_{\mu\nu\rho\sigma}R^{\mu\nu\rho\sigma}\) etc., pure gauge term e.g. $F^4$ , mixed terms such as $RF^2$ and many other terms. That makes the construction quite involved. A superconformal approach was taken in \cite{hanaki2007supersymmetric} to develop an off-shell formulation involving  the  Weyl multiplet which is locally gauge invariant under the superconformal group. The resulting conformal supergravity is reduced to Poincar$\text{e}^\prime$ supergravity by introducing conformal compensator in the hypermultiplet sector and introducing expectation values for some of the fields. Thus they obtain the supersymmetric completion of the \(A \wedge Tr R \wedge R\) term in   \( {\mathcal N}=2 \) supergravity.

The basic construction  conformal supergravity (Weyl multiplet), coupled to $n_v+1$ conformal vector multiplets and a single compensator hypermultiplet
\ben {\mathcal L} = {\mathcal L}_0 + {\mathcal L}_1 = {\mathcal L}_0^V + {\mathcal L}_0^V + {\mathcal L}_1  , \ee
where ${\mathcal L}_0$ contains terms quadratic in derivatives, which may split into vector multiplets ${\mathcal L}_0^V$  and hypermultiplet ${\mathcal L}_0^V$, while \({\mathcal L}_1\) contains terms quartic in derivatives. The full Lagrangian              ${\mathcal L}$ contains auxiliary fields. In \cite{Cremonini:2008tw}, the auxiliary fields are integrated out using their equations of motion to obtain the on-shell two-derivative, which with a few redefinitions can be brought down to the standard two-derivative ${\mathcal N}=2$ supergravity action coupled to $n_v$ vector multiplets. Finally it is truncated to pure supergravity by setting the scalars to appropriate constants and defining a single graviphoton it reduces to the conventional on-shell supergravity Lagrangian, as given in \eqref{1charge-action}.

In order to obtain the Lagrangian upto four-derivative, consisting of only the physical fields one can integrate the auxiliary fields out from the Lagrangian ${\mathcal L} = {\mathcal L}_0 + {\mathcal L}_1$. It turns out that the lowest order expressions that was obtained in the two-derivative case are sufficient if we are interested in the terms upto linear order in $c_{2I}$. However it needs to be noted the values of the scalar fields will be modified by terms linear in $c_{2I}$ due to modified very special geometry constraint. The contribution coming from ${\mathcal L}_0$ turns out to be 
\be
 e^{-1} \mathcal{L}_0 =-R-\frac{3}{4}F^2 + \frac{1}{4}  \epsilon^{\mu \nu \rho \lambda \sigma} A_\mu F_{\nu \rho} F_{\lambda \sigma} +12 g^2 + \frac{c_2}{24} \Bigg( \frac{1}{16} R F^2 +\frac{1}{64} (F^2)^2 - \frac{5}{4} g^2 F^2 \Bigg) 
 \ee
The contribution from the quartic order Lagrangians are given by as written below segregated in the gauged and ungauged part,
\be \begin{split}
 e^{-1} \mathcal{L}_1^{\text{gauged}} & = -\frac{1}{6} c_2 g^2  \epsilon^{\mu \nu \rho \lambda \sigma} A_\mu F_{\nu \rho} F_{\lambda \sigma} \\
 e^{-1} \mathcal{L}_1^{\text{ungauged}} &= \frac{c_2}{24} \left(  \frac{1}{16} \epsilon_{\mu \nu \rho \lambda \sigma} A^\mu R^{\nu \rho \delta \gamma}{R^{\lambda \sigma}}_{\delta \gamma} + \frac{1}{8}C^2_{\mu \nu \rho \sigma} + \frac{3}{16} C_{\mu \nu \rho \lambda}F^{\mu \nu} F^{\rho \lambda} -  F^{\mu \rho}F_{\rho \nu} R^\nu_\mu + \right. \\
 &-\frac{1}{8} RF^2 + \frac{3}{2} F_{\mu \nu} \nabla^\nu \nabla_\rho F^{\mu \rho} + \frac{3}{4}  \nabla^\mu F^{\nu \rho} \nabla_\mu F_{\nu \rho} + \frac{3}{4}  \nabla^\mu F^{\nu \rho} \nabla_\nu F_{\rho \mu}+\\
 &+\frac{1}{8} \epsilon_{\mu \nu \rho \lambda \sigma} F^{\mu \nu}(3 F^{\rho \lambda} \nabla_\delta F^{\sigma \delta} + 4 F^{\rho \delta} \nabla_\delta F^{\lambda \sigma} + 6 {F^\rho}_\delta \nabla^\lambda F^{\sigma \delta})\\
 &\left. \frac{45}{64} F_{\mu \nu }F^{\nu \rho} F_{\rho \lambda} F^{\lambda \mu} - \frac{45}{256} (F^2)^2 \right).       \label{eq:36A}
\end{split} 
\ee

Making a redefinition of $A_\mu$ to write the kinetic term in a canonical form,
\be
A_\mu^{\text{new}} = \sqrt{3}\left( 1 + \frac{5}{144} c_2 g^2 \right) A_\mu^{\text{old}} ,
\ee
we obtain the Lagrangian  upto four-derivative terms, as follows
\begin{equation}
 \begin{aligned}
 e^{-1} \mathcal{L} & =-R-\frac{1}{4}F^2 + \frac{1}{12 \sqrt{3}} \Bigg( 1-\frac{1}{6} c_2 g^2\Bigg) \epsilon^{\mu \nu \rho \lambda \sigma} A_\mu F_{\nu \rho} F_{\lambda \sigma} +12 g^2 + \frac{c_2}{24} \Bigg( \frac{1}{48} R F^2 +\frac{1}{576} (F^2)^2 \Bigg) +\\
 & + \frac{c_2}{24} \left(  \frac{1}{16 \sqrt{3}} \epsilon_{\mu \nu \rho \lambda \sigma} A^\mu R^{\nu \rho \delta \gamma}{R^{\lambda \sigma}}_{\delta \gamma} + \frac{1}{8}C^2_{\mu \nu \rho \sigma}+\frac{1}{16}C_{\mu \nu \rho \lambda}F^{\mu \nu} F^{\rho \lambda} - \frac{1}{3} F^{\mu \rho}F_{\rho \nu} R^\nu_\mu + \right. \\
 &-\frac{1}{24} RF^2 + \frac{1}{2} F_{\mu \nu} \nabla^\nu \nabla_\rho F^{\mu \rho}+ \frac{1}{4}  \nabla^\mu F^{\nu \rho} \nabla_\mu F_{\nu \rho}+\frac{1}{4}  \nabla^\mu F^{\nu \rho} \nabla_\nu F_{\rho \mu}+\\
 &+\frac{1}{32 \sqrt{3}} \epsilon_{\mu \nu \rho \lambda \sigma} F^{\mu \nu}(3 F^{\rho \lambda} \nabla_\delta F^{\sigma \delta}+4 F^{\rho \delta} \nabla_\delta F^{\lambda \sigma} +6 {F^\rho}_\delta \nabla^\lambda F^{\sigma \delta})\\
 &\left.  \frac{5}{64} F_{\mu \nu }F^{\nu \rho} F_{\rho \lambda} F^{\lambda \mu} - \frac{5}{256} (F^2)^2  \right)      \label{eq:36}
\end{aligned} 
\end{equation}
This represents the higher derivative corrected version of the one-charge quadratic action given in \eqref{eq:1},

A similar four-derivative correction of the black hole solution with metric and the gauge field given in  \eqref{1-charge-bh}  is obtained in \cite{Cremonini:2008tw} perturbatively up to first order in the parameter \(c_2\), and are given as follows:
\begin{equation}
\begin{aligned}
&ds^2=  \frac {f}{H^2} dt^2 - H \left( \frac{1}{f}dr^2 + r^2 d x^2+ r^2 d y^2+r^2 d z^2   \right)  \\
& f = k - \frac{\mu}{r^2 }+ g^2 r^2 H^3 + c_2 f_1(r)  , \quad H=1+\frac{Q}{r^2} + c_2 h_1  \\
& A = \sqrt{\frac{3(kQ+\mu)}{Q}} \Bigg(  1- \frac{1+ c_2 a_1}{H}\Bigg) dt    \label{eq:37},
\end{aligned}  
\end{equation}
where ,
\begin{equation}
\begin{split}
h_1 &= - \frac{Q(kQ+\mu)}{72 r^6 H_0} ,\quad
f_1= \frac{-5 g^2 Q(kQ +\mu)}{72 r^4} + \frac{\mu^2}{96 r^6 H_0} \\
a_1 &= \frac{Q}{144 r^6 H_0^3}(4(kQ+ \mu)-3\mu -\frac{3Q\mu}{r^2})             \label{eq:38}
\end{split}
\end{equation}

We are interested for \(k=0\), for which the above equations\eqref{eq:38} transform into, 
\begin{equation}
\begin{split}
&H= H_0 + \frac{c_2}{24} \Bigg( \frac{-Q \mu}{3 r^6 H_0^2}\Bigg) , \quad
f=H_0+\frac{c_2}{24}\Bigg( \frac{-8 g^2 \mu Q)}{3 r^4} + \frac{\mu^2}{4 r^6 H_0} \Bigg) , \\
&A_t= A_{t0}+\frac{c_2}{24}\Bigg( \frac{\sqrt{3 Q \mu}}{2 r^8 H_0^4}(\mu r^2- Q \mu)\Bigg)      \label{eq:39}
\end{split}
\end{equation}
In the above equations \eqref{eq:39}, the functions \(f_0\),  \(H_0 \) and \(A_{t0}\) are identical to those defined in equations \eqref{1-charge-bh}, %while \(H_0(r) = 1 + \frac{Q}{r^2}\).  

The equations of motion are given as follows \cite{Kats:2006xp}\cite{Liu:2016hqb}. The general Maxwell equation, with the higher derivative correction will be
\begin{equation}
\begin{aligned}
&\nabla_\nu F^{\nu \mu}  + \epsilon^{\mu \beta \rho \lambda \sigma}(\alpha F_{\beta \rho} F_{ \lambda \sigma}  +  \lambda {R^\delta}_{\eta \beta \rho}  {R^\eta}_{\delta \lambda \sigma} ) + 4 b_4 \nabla_\nu(R F^{\mu \nu}) + \\
&+ 2 b_5 \nabla_\nu (R^{\mu \rho} {F_\rho}^\nu-R^{\nu \rho} {F_\rho}^\mu) + 4 b_6 \nabla_\nu(R^{\alpha \beta \mu \nu} F_{\alpha \beta})+ 8 b_7 \nabla_\nu(F_{\rho \sigma} F^{\rho \sigma} F^{\mu \nu})+ \\
&- 4 b_8 \nabla_\nu \Box F^{\mu \nu} - 2 b_9 \nabla_\nu \nabla_\rho(\nabla^\mu F^{\rho \nu}-\nabla^\nu F^{\rho \mu}) =0  .      \label{eq:40}
\end{aligned}
\end{equation}
The Einstein equation can be written in the  following form,
\begin{equation}
-R_{\mu \nu}+ 4g^2 g_{\mu \nu} +\frac{1}{12} g_{\mu \nu} F^2 -\frac{1}{2} F_{\mu \lambda} {F_\nu}^\lambda =T_{\mu \nu} -\frac{1}{3} g_{\mu \nu} T .   \label{eq:41}
\end{equation}
where all the higher derivative corrections are assembled into a form of energy momentum tensor \(T_{\mu\nu}\) and \(T\) represent the trace of it. This energy momentum tensor \(T_{\mu\nu}\) is given by
\begin{equation}
\begin{aligned}
T_{\mu \nu}&= b_1(g_{\mu \nu} R^2 - 4 R R_{\mu \nu}+ 4 \nabla_\nu \nabla_\mu R - 4 g_{\mu \nu}  \Box R)+\\
&+b_2(g_{\mu \nu}  R_{\rho \sigma}R^{\rho \sigma}+ 4 \nabla_\alpha \nabla_\nu R^\alpha_\mu -2 \Box R_{\mu \nu}- g_{\mu \nu} \Box R-4R^\alpha_\mu R_{\alpha \nu})+\\
&+b_3(g_{\mu \nu} R_{\alpha \beta \gamma \delta}R^{\alpha \beta \gamma \delta} - 4 R_{\mu \alpha \beta \gamma} {R_\nu}^{\alpha \beta \gamma}- 8 \Box R_{\mu \nu}+4 \nabla_\nu \nabla_\mu R +8 R^\alpha_\mu R_{\alpha \nu}- 8 R^{\alpha \beta}R_{\mu \alpha \nu \beta})+\\
&+b_4(g_{\mu \nu} RF^2- 4RF_\mu^\sigma F_{\nu \sigma} -2 F^2 R_{\mu \nu} + 2 \nabla_\mu \nabla_\nu F^2 -2 g_{\mu \nu} \Box F^2)+\\
&+b_5(g_{\mu \nu} R^{\kappa \lambda}F_{\kappa \rho}{F_\lambda}^\rho - 4 R_{\nu \sigma} F_{\mu \rho}F^{\sigma \rho}-2R^{\alpha \beta}F_{\alpha \mu} F_{\beta \nu}- g_{\mu \nu} \nabla_\alpha \nabla_\beta({F^\alpha}_\rho F^{\beta \rho})+\\
&+2 \nabla_\alpha \nabla_\nu(F_{\mu \beta} F^{\alpha \beta}) -\Box(F_{\mu \rho}{F_\nu}^\rho))+\\
&+b_6(g_{\mu \nu} R^{\kappa \lambda \rho \sigma} F_{\kappa \lambda}F_{\rho \sigma}-6 F_{\alpha \nu} F^{\beta \gamma} {R^\alpha}_{\mu \beta \gamma}-4 \nabla_\beta \nabla_\alpha({F^\alpha}_\mu {F^\beta}_\nu))+\\
&+b_7(g_{\mu \nu}( F^2)^2-8 F^2 {F_\mu}^\sigma F_{\nu \sigma})+\\
&+b_8(g_{\mu \nu}(\nabla_\kappa F_{\rho \sigma})(\nabla^\kappa F^{\rho \sigma})-2(\nabla_\mu F_{\alpha \beta})(\nabla_\nu F^{\alpha \beta})-4 (\nabla_\alpha F_{\beta \mu})(\nabla^\alpha {F^\beta}_\nu)+\\
&+4\nabla_\alpha(F_{\nu \beta} \nabla^\alpha {F_\mu}^\beta)+4\nabla_\alpha (F_{\nu \beta} \nabla_\mu F^{\alpha\beta})- 4\nabla_\alpha({F^\alpha}_\beta \nabla_\nu {F_\mu}^\beta))+\\
&+b_9 (g_{\mu \nu}(\nabla_\kappa F_{\rho \sigma})(\nabla^\rho F^{\kappa \sigma})-4(\nabla_\mu F^{\alpha \beta})(\nabla_\alpha F_{\nu \beta})-2(\nabla_\alpha F_{\beta \mu})(\nabla^\beta {F^\alpha}_\nu)+\\
&+ 2 \nabla_\alpha(F_{\nu \beta} \nabla^\alpha {F_\mu}^\beta)+2 \nabla_\alpha(F_{\nu \beta} \nabla_\mu F^{\alpha \beta})-2 \nabla_\alpha({F^\alpha}_\beta \nabla_\nu {F_\mu}^\beta))
\end{aligned}
\end{equation}
The parameters used in the above equations, \eqref{eq:40} and \eqref{eq:41}, can be expressed in terms of \(c_2\) and \(g\) as follows,
 \begin{equation}
 \begin{aligned}
&\alpha =\left(\frac {1}{4\sqrt{3}}\right)\left(1-\frac{c_2 g^2}{6}\right),   \quad \quad \quad \quad \quad \quad \lambda =  \left(\frac {c_2}{24}\right) \left(\frac {1}{16 \sqrt{3}}\right), \\
&b_1 = \left(\frac {c_2}{192}\right) \left(\frac{1}{6}\right), \quad \quad b_2 =-\left(\frac {c_2}{192}\right) \left(\frac{4}{3}\right), \quad\quad b_3 = \left(\frac {c_2}{192}\right), \\
&b_4 = -\left(\frac {c_2}{192}\right) \left(\frac{1}{4}\right), \quad \quad b_5 =\left(\frac {c_2}{192}\right) \left(\frac{10}{3}\right), \quad\quad b_6 =- \left(\frac {c_2}{192}\right)\left(\frac{1}{2}\right), \\
&b_7 = \left(\frac {c_2}{192}\right) \left(\frac{49}{288}\right), \quad \quad b_8 =2 \left(\frac {c_2}{192}\right),\quad \quad \quad\quad b_9 =-2  \left(\frac {c_2}{192}\right)
 \end{aligned}
 \end{equation}
 
These constitutes the essential holographic setup for our analysis. Building on this framework, we will make an attempt to analyze the stability of its near-horizon limit. We will apply perturbations to the above equations, employing the same perturbative approach that was utilized in the quadratic order analysis in sections \ref{sec:2}  and \ref{sec:3}, and explore the effects of such corrections.
 
\subsection{Near horizon analysis}

Continuing our analysis of the black hole solution~\eqref{eq:37}, we observe that, even with the inclusion of higher-order correction terms in the zero-temperature limit, the geometry remains characterized by an \( \mathrm{AdS}_2 \times \mathbb{R}^3 \) configuration,

\begin{equation}
ds^2 = \frac{L^2}{\rho^2} (d \hat{t}^2- d\rho^2)- d \hat{x}^2-d \hat{y}^2-d \hat{z}^2  \label{eq:42}
 \end{equation}
 
 However, both the AdS radius \(L^2\)  and the gauge field \(A_t \) undergo modifications in the near-horizon region, as given below,

 \begin{equation}
L^2 = \frac{1}{12g^2} - \frac{17 c_2}{3456}  \hspace{0.05cm} ,        \quad \quad  \quad A_t = -\frac{1}{\rho} \left(\frac{1}{\sqrt{6} g }+\frac{47 c_2 g}{576 \sqrt{6}}\right)    \label{eq:43}
 \end{equation}

where, as earlier \( \rho = \frac{1}{r - r_h}\). We plug in the perturbative ansatz defined in equations~\eqref{perturbation}, in the equation of motion with the near-horizon geometry of the black hole as a background, and obtained linearised form of the equations. These lead to a set of four coupled differential equations. The Maxwells equation \eqref{eq:40} gives rise to the following equation,  

\begin{equation}
\begin{aligned}
12& g^2 \rho^2(a_i '' -{\partial_t}^2 a_i) -k^2 a_i -16  \sqrt{6} i g k \alpha \epsilon^{i j} a_j  - 24 \sqrt{6} g^3 \rho^2 (h_{ti}' + 8 \sqrt{6} i g k \lambda  \epsilon^{ij} h_{tj}')+ \\
&+ \frac{1}{144} c_2 ( - 3 (11 g^2 k^2 + k^4)  a_i + 4 i \sqrt{6} g k \epsilon^{ij} (55 g^2 \alpha_0 - 576 \alpha_1) a_j + \\
&+ 6 g^2 \rho ^2 ( ( 12 k^2 - 109 g^2 )( a_i'' - \partial_t^2 a_i)+ 288 g^2 \rho (\partial_t^2 (a_i' )-a_i^{(3)})+\\
&- 72 g^2 \rho^2 ( a_i^{(4)}+ \partial_t^4 a_i)+ 144 g^2 \rho^2  \partial_t^2  a_i'' +\\
& +\sqrt{6} g (137 g^2 - 2 k^2)  h_{ti}' +   192 i g^2 k \epsilon^{ij} (17 g^2 \lambda_0 - 144 \lambda_1)  h_{tj}' +\\
&  +96 \sqrt{6} g^3 \rho  h_{ti}''  + 24 \sqrt{6} g^3 \rho^2 (h_{ti}^{(3)}-\partial_t^2  h_{ti}' ))=0      \label{eq:44}
\end{aligned}
\end{equation}

The Einstein equation \eqref{eq:41} gives rise to the following three equations.
 \begin{equation}
\begin{aligned}
 & -6 g^2 \rho^2 h_{ti}'' -12 g^2 \rho h_{ti}'+\frac{k^2}{2} h_{ti}+\frac{i k}{2} \partial_t h_{xi} + \sqrt{6} g a_i'  +\\
 &\quad+ i 4 \sqrt{6} k \lambda \epsilon^{ij}(12 g^3 \rho^2 h_{tj}''+24 g^3 \rho h_{tj}'- k^2 g  h_{tj}-ik g \partial_t h_{xj}+2\sqrt{6} g^2 a_j')  +\\
 & \frac{1}{576} c_2 \Bigg[ -2 \left(163 g^2 k^2 + 2 k^4\right) h_{ti}
- 1008 i \sqrt{6} g^5 k \rho \lambda_0 \epsilon^{ij} \left(2 h_{tj}' + \rho h_{tj}''\right)+ \\
& - 4 i k^3 \partial_t h_{xi}
+ 8 \sqrt{6} g k^2 \left( a_i' + 288 \lambda_1 \epsilon^{ij} \left(-i k h_{tj} + \partial_t h_{xj} \right) \right) +\\
& + \sqrt{6} g^3 \Big(
-411 a_i'
+ 4 \Big[
55 i k^3 \lambda_0 \epsilon^{ij} h_{tj}
+ 13824 i k \rho \lambda_1 \epsilon^{ij} h_{tj}'
- 48 \rho a_i''  +\\
& \quad + 6912 i k \rho^2 \lambda_1 \epsilon^{ij} h_{tj}''
- 24 \rho^2 a_i^{(3)}
- 55 k^2 \lambda_0 \epsilon^{ij} \partial_t h_{xj}
+ 48 \rho \partial_t^2 a_i
+ 24 \rho^2 \partial_t^2 a_i'
\Big]
\Big) +\\
& + 36 g^4 \Big(
-136 i k \lambda_0 \epsilon^{ij} a_j'
+ \rho \Big[
174 h_{ti}'
+ \rho \Big(
-105 h_{ti}''
+ 16 \rho \Big(
-8 h_{ti}^{(3)}
- \rho h_{ti}^{(4)} + 4 \partial_t^2 h_{ti}'
+ \rho \partial_t^2 h_{ti}''
\Big)
\Big)
\Big]
\Big)+ \\
& + 2 g^2 k \Big(
13824 i \lambda_1 \epsilon^{ij} a_j'
- 163 i \partial_t h_{xi}
+ 24 \rho \Big(
4 k h_{ti}'
+ 2 k \rho h_{ti}''
+ i \rho \Big(
\partial_t h_{xi}''
+ i k \partial_t^2 h_{ti} +\\
& \quad - \partial_t^3 h_{xi}
\Big)
\Big)
\Big)
\Bigg] = 0     \label{eq:45}
\end{aligned}
\end{equation}

 \begin{equation}
 \begin{aligned}
 & h_{xi}'' - {\partial_t}^2 h_{xi} +i k \partial_t h_{ti}-i 8 \sqrt{6}g k \lambda  \epsilon^{ij} (h_{xj}''-{\partial_t}^2 h_{xj}+ik\partial_t h_{tj})+\\
&+ \frac{1}{288} c_2 \Bigg[
  \Big(39 g^2 - 4 k^2\Big) h_{xi}''
  + 4 k^2 \Big(-i k\partial _t h_{ti} + \partial _t^2 h_{xi} \Big)
  - 84 \sqrt{6} g^3 k \lambda_0 \Big(-i h_{xj}'' + k \partial_t h_{tj} + i h_{xj}'' \Big)+ \\
 & + 2304 \sqrt{6} g k \lambda_1 \Big(-i h_{xj}'' + k \partial_t h_{tj} + i \partial _t^2 h_{xj} \Big)+\\
 &+ 3 g^2 \Big[13 i k \partial_t h_{ti} - 13 \partial _t^2 h_{xi} + 16 \rho \Big(4 h_{xi}^{(3)} + 4 i k \partial_t h_{ti}' - 4 \partial _t^2 h_{xi}'+\\
 & + \rho \Big( {h_{xi}}^{(4)} - 2 \partial_t^2 {h_{xi}''} + i k \Big( \partial_t h_{ti}'' - \partial_t^3 h_{ti}  \Big) + \partial _t^4 h_{xi} \Big) \Big) \Big] \Bigg] = 0       \label{eq:46}
\end{aligned}
\end{equation}
 
\begin{equation}
\begin{aligned}
&\sqrt{6} i g \partial_t a_i-6 i g^2 \rho^2 \partial_t {h_{ti}}' -\frac{k}{2} h_{xi}'+i 4 \sqrt{6} k \epsilon^{ij}  \lambda \epsilon^{ij}(k h_{xj}'+12 g^3 \rho^2 \partial_t h_{tj}'+2 \sqrt{6}  g^2 i \partial_t a_j)+\\
&+\frac{1}{576} c_2 \Bigg[\Big(309 g^2 k + 4 k^3 \Big) h_{xi}' + g \Bigg( 8 i \sqrt{6} k^2 \Big( 288 \epsilon^{ij}  \lambda_1 h_{xj}'+ \partial_t a_i \Big) - 624 \sqrt{6} g^4 k \rho^2 \lambda_0 \epsilon^{ij}  \partial_t h_{tj}' + \\
&+ 48 g k \Big( -576 \lambda_1  \epsilon^{ij}  \partial_t a_j + \rho ( -2 h_{xi}'' - \rho h_{xi}^{(3)} - 2 i k \partial_t h_{ti} + i k \rho \partial_t h_{ti}' + 2 \partial_t ^2 h_{xi}+ \rho \partial_t^2 h_{xi}' ) \Big) \Bigg)+\\
&- i \sqrt{6} g^2 \Big( 84 k^2 \lambda_0 \epsilon^{ij} h_{xj}' + 377 \partial_t a_i + 96 \rho^2 \left( -288 i k \lambda_1 \epsilon^{ij}  \partial_t h_{tj}'+ \partial_t a_i'' - \partial_t^3 a_i \right) \Big)+ \\
&+ 48 g^3 \Bigg( 68 k \lambda_0 \epsilon^{ij}  \partial_t a_j - i \rho^2 \left( -61 \partial_t h_{ti}' + 12 \rho ( 4 \partial_t h_{ti}'' + \rho \partial_t h_{ti}^{(3)} - \rho \partial_t^3 h_{ti}' ) \right) \Bigg) \Bigg] =0       \label{eq:47}
\end{aligned}
\end{equation}

As one can observe these differential equations involve third and fourth order of derivatives unlike its quadratic counterpart. These higher order derivatives can be traced back to our Lagrangian \eqref{eq:36}, which itself is quartic in derivatives. Therefore it may be inflicted with instabilities associated with higher derivatives as predicted by Ostrogradsky's theorem \footnote{Ostrogradsky's theorem, which involves construction of Hamiltonian formalism for non-degenerate higher derivative shows that it does not have energy bounded from below and so are unstable as shown by  \cite{Ostrogradsky:1850fid}. A nice review about Ostrogradsky theorem for mechanical system is given in \cite{Woodard:2015zca}. A number of works that discuss the possibility of evading this instability, such as \cite{Klein:2016aiq} discussed how to exercise the Ostrogradsky ghosts in coupled system, \cite{Motohashi:2014opa} discussed theories with third order time derivatives, \cite{Crisostomi:2017aim} discusses degeneracy condition for higher field theories.
A discussion of higher derivative theories of gravity has appeared in \cite{Wooliams:2013sa}, where they considered the action given by linear superposition of all permissible quartic terms in graviational theory,
\be
S = - \int d^4x \sqrt{-g} \left( \alpha R_{\mu\nu} R^{\mu\nu} - \beta R^2 + \gamma \kappa^{-2} R\right),
\ee
where $\kappa^2=32\pi G$ and which can be written in terms of Weyl tensor as well and they found, among other things,  $\alpha \neq 0$ leads to Ostrogradsky instability. }.

This issue of instability due to higher derivative in pure gravity theory has been addressed in  \cite{Boulware:1983td}. They considered conformally invariant Weyl action, which is the integral of the square of the Weyl tensor and can be considered to be a candidate for quantum gravity. If dynamical breakdown of scale invariance occurs in quantum field theory, one expects that at long distance the conformal invariance will be broken leaving general coordinate invariance. The resultant theory is described by an effective action with Hilbert action playing the leading term. In our higher derivative action the pure gravity part of the Lagrangian \eqref{eq:36} is exactly conformally invariant Weyl action and solution of the classical theory runs the risk of exhibiting instability through the solutions of the linearised equations with arbitrarily large negative energy. 

It has been argued\cite{Boulware:1983td}, that asymptotically flat solutions of the classical equations all have zero energy in the following way. Since the action is quartic, classical potential grows linearly with distance, like Yang-Mills theory where only systems with zero color have finite energy. Because of the linear dependence on distance, if total color is not zero, there will be a color field extending to infinity leading to divergent energy. So color is confined in Yang-Mills theory and energy is confined in this gravitational theory. However, if we add an Einstein term, being lower order in derivative, that will dominate the large distance dynamics and long range potential will become Coulombic. So there will be no confinement and energy can have either sign, making the system susceptible to negative energy problem. The mathematical analysis using canonical formulation establishing this claim is reviewed in the Appendix.

In our case, it is five dimensional and a similar calculation would lead to a logarithmic dependence on distance of the potential energy. A  similar argument may work though in our case we are interested in asymptotically AdS solution and the dominance of Einstein term at large distance may make it unstable. It requires a full fledged canonical analysis along the line of \cite{Boulware:1983td}. The necessary formalism has already been discussed in \cite{Boulware:1984ch}.

\section{Dual field theory} 
 \label{sec:5}
The conjecture of gauge/gravity duality establishes connection between the weakly coupled gravitational theories and strongly coupled gauge theories\cite{Maldacena:1997re} and vice versa. It relates Type IIB string theory on \(AdS_5 \times Y^5\) to \(\mathcal{N}=4\) super Yang Mills (SYM) with gauge group \(SU(N)\). The AdS/CFT dictionary relates the 't Hooft coupling $\lambda = g_{YM}^2 N$ as
\be \frac{L^4}{\alpha^{\prime 2}} = 4 \pi g_s N = g_{YM}^2 N \ee
where \( L\) is the radius of AdS\(_5\). In terms of the supergravity, we have  \(\mathcal{N}=8\) Type IIB supergravity on AdS\(_5\), related to an \(\mathcal{N}=4\) Super Yang-Mills theory with \(SO(6)\) R-symmetry living on the world volume of  N D3 branes\cite{Cvetic:1999xp} \cite{Cremonini:2008tw}.
The present case corresponds to compactification on \(AdS_5 \times Y^5\), where $Y^5$ is a Sasaki-Einstein manifold \(\mathcal{N}=2\) gauged supergravity theory with a \(U(1) \times U(1) \times U(1)\) gauge group \cite{Cremonini:2008tw}. This corresponds to the decoupling limit of rotating D3-branes and the dual theory is identified as an \(\mathcal{N}=1\) Super Yang-Mills theory in (3+1)-dimensions.

Black holes in Einstein–Maxwell–Anti-de Sitter system was studied in \cite{Chamblin:1999tk}, which arises as near horizon limit rotating D3 branes. The gravity theory is thus regarded as the effective theory of the strongly coupled field theory living on the rotating brane's world volume. So it sheds lights on the thermodynamics of the dual superconformal theory with background global current switched on. In \cite{erdmenger2015hydrodynamic} they considered a stack of N spinning D3-branes in flat space, assuming all spins are equal and examine the effective hydrodynamics on the world-volume of the spinning D3 branes. In \cite{kraus1999coulomb} they studied the Coulomb branch of ${\mathcal N}=4$ SYM using the supergravity description as rotating D3 branes.
A direct analysis of the field theory dual to the spinning D3 branes inside this instability regime of parameter in particular, the behaviour of the spatially modulated system may be interesting to study.  

Most studies focus on phenomena involving quadratic-order terms, largely due to the technical challenges associated with higher-derivative corrections. An extension of this analysis beyond quadratic terms to higher order correction terms should correspond to subleading corrections in the 't Hooft coupling on the field theory side\cite{Hanada:2008ez}. Including such corrections is crucial for capturing subleading effects, thereby refining the holographic models.

\section{Conclusion}
\label{sec:6}

In this article we explore the possible instabilities in a top-down approach for a gravitational system obtained in the context of string theory/supergravity theory. We consider a one-charge version of  ${\mathcal N}=2$, D=5 supergravity, which can be obtained from $S^5$ reduction of type IIB supergravity. The Lagrangian of this one charge version involves Einstein-Hilbert action coupled with the gauge Chern-Simons term as well as the mixed gauge-gravitational Chern-Simons term, with specific coefficients in a five-dimensional spacetime \cite{Cvetic:1999xp} and analysed the instabilities of an asymptotically AdS charged black hole solution from a holographic perspective.

To begin with we  turned our attention to the near-horizon region, which for the present case reduces to an $AdS_2 \times \mathbb{R}^3$ and consider instability due to the gauge Chern-Simons term only. We find the instability depends on the strength of the Chern-Simons term (\( \alpha \)). There exists a critical value of the strength , beyond which the BF bound is violated and the instability sets in. This critical value matches exactly with the coefficients of Chern-Simons term in the action obtained in the top down approach implying the solution is marginally stable. It may be interesting to understand whether this precise agreement has any relation with the supersymmetry.  On the other hand turning on only the gauge-gravitational Chern-Simons term alone does not lead to a violation of the BF bound and does not generate instability on its own.

In contrast, when both gauge Chern-Simons and mixed gauge-gravitational Chern-Simons terms are simultaneously turned on, even with a small coefficient for the gauge-gravitational term, the BF bound is violated. This signals the appearance of an unstable mode in the bulk. The system thus undergoes a transition from a stable to an unstable phase depending on the interplay between the gauge and gauge-gravitational Chern-Simons terms.
Moreover, as the coefficient of the gauge-gravitational Chern-Simons term increases, the violation of the BF bound becomes more significant, indicating stronger instability within the bulk geometry. Analysis of the normal modes of the full fledged black hole solution also exhibits similar behaviors.  

Since gauge-gravitational Chern-Simons terms involves higher order derivatives, we should consider the solution up to the quartic order of derivatives. Primarily we used the argument that the gauge-gravitational Chern-Simons can be treated non-perturbatively unlike the other higher derivative terms. We included higher-derivative corrections up to fourth order in derivatives \cite{Cremonini:2008tw}, corresponding to the leading-order correction. Employing certain  perturbative ansatz, we carried out a near-horizon analysis within this extended framework. Perhaps, in accordance with the preliminary results, if the correction terms are considered negligibly small, such instabilities may arise in the near-horizon region with more refined precisions.  However, such higher derivative equations may be inflicted with Ostrogradsky's singularity and that requires an analysis using the canonical formulation. We briefly mentioned the way in which this analysis needs to be done and  briefly reviewed a similar analysis of a particular higher gravity theory. If such instabilities are observed, it would motivate a full fledged analysis of the coupled bulk equations to uncover the corresponding spatially modulated instabilities in the full gravitational background.

A natural continuation of the present work is to consider the rigorous analysis of the higher derivative theory in the canonical formulation along the line of \cite{Boulware:1983td} for which the necessary formalism has already been discussed in \cite{Boulware:1984ch}. Since we have already identified the instabilities, a natural sequel is to identify its final configuration along the line of \cite{Ooguri:2010kt,Donos:2012wi}. Similar reductions of string theory may lead to supergravity in four dimensions\cite{Cvetic:1999xp}. Though four dimensional theories do not admit Chern-Simons term but it may involve pseudoscalars. As shown in \cite{Donos:2011bh}, four dimensional theories with pseudoscalar fields may lead to instabilities, which can be explored in similar top down models. A very interesting extension would be to explore the affects of the instability stemming from the gauge and mixed gauge-gravitational Chern-Simons term in the dual field theory along the line of \cite{Landsteiner:2011iq}.  We hope to return with some of these issues in future.

%%%%%%%%%%%%%%%%%%%%%%%%%%%%

\appendix
\section{Energy for scale-invariant quartic Lagrangian}
\label{sec:7}

In \cite{Boulware:1983td} they considered a general scale invariant gravitational action and established that for space-time being asymptotically flat, in a certain sense the energy in a theory with linearised perturbation is zero. In what follows, we will briefly describe their analysis. The action is given by
\be S = - \frac{1}{4} \int d^4x \sqrt{-g} \left( \alpha c^{\mu\nu\lambda\sigma} c_{\mu\nu\lambda\sigma} + \beta R^2\right). \ee

Since it involves higher derivative a  canonical formulation of the scale invariant gravity theory \cite{Boulware:1984ch} is required. For a space-like surface $\Sigma$ in an asymptotically flat space-time $(M, g_{\mu\nu})$, the canonical variables, consists of conjugate pairs, namely, a usual three-metric \(g_{ij}\) and its conjugate momenta \(p^{ij}\), (which involves third order time derivative of the metric) and extra degrees of freedom \( Q^{ij} \) and its conjugate \(P^{ij} \) ( which occurs due to the higher derivative terms), they are given by \( Q^{ij} = \sqrt{g} \left( 2 \alpha C^{0 i ~ j }_{~~0} + \beta g^{ij} R \right) \), \(P^{ij} = 2 K_{ij}\),  \(K_{ij}\) represents the extrinsic curvature. 

Its invariance under the diffeomorphism group is reflected in the presence of the  constraints. They are time-time component and time-space components of classical field equations,
\be \begin{split} \label{constraint}
 C &= \frac{1}{2} p^{ij}P_{ij} - \alpha \sqrt{g} C^{0ijk}C^0_{~ijk} - \frac{1}{2} \alpha \sqrt{g} Q^{Tij}Q^T_{~ij} - \frac{1}{36} \beta \sqrt{g} Q^2 \\ & - \frac{1}{4}  Q^{ij}P_{ij} P - \frac{1}{8} Q(P^{ij}P_{ij} -P^2) 
 + \frac{1}{2}  \hphantom{R}\!\!\!^3 R Q - \hphantom{R}\!\!\!^3R_{ij}Q^{ij}  - D_iD_j Q^{ij} = 0,\\
 C_k &= \frac{1}{2} Q^{ij} D_k P_{ij} - D_i (P_{jk} Q^{ij}) + D_i P_k^{~i}
 \end{split}
\ee
where \(D_i, \hphantom{R}\!\!\!^3 R_{ij}, \hphantom{R}\!\!\!^3 R \) are covariant derivative, Ricci tensor and scalar curvature of \( g_{ij}\), \(P\) is the trace of \(P_{ij}\) and \(C^0_{~ijk}\) can be expressed in terms of spatial derivative of \(P_{ij}\).

 The generators of the diffeomorphism are given by volume integrals of the constraints up to surface terms, which are fixed by the condition that the generators are differentiable functions on the phase space, \(\Gamma = \{ g, p, Q,P\}\). 
 The generating functions are 
 \be H_N = \int N(x) C(x) d^3x + \int N D_j Q^{ij} dS_i\\
 P_{N^k} = \int N^k(x) C_k(x) d^3x + \int N^k \{ p_k^{~i} - Q^{ij}P_{jk}\} dS_i,
 \ee
 where\(N\) and \( N_k\) are asymptotically well behaved functions and vector fields on \(\Sigma\).
 
 The total energy \(E\) is the Hamiltonian when the constraints are satisfied, which is given along with the momentum  as
 \be E=\int D_j Q^{ij} dS_i , \quad P_k = - \int \{ P_k^{~i} - Q^{ij}P_{jk} \} dS_i , \ee
Using the constraint \( C=0\), this becomes
\be \begin{split} \label{energy}
 E &= \int \frac{1}{2} p^{ij}P_{ij} - \alpha \sqrt{g} C^{0ijk}C^0_{~ijk} - \frac{1}{2} \alpha \sqrt{g} Q^{Tij}Q^T_{~ij} - \frac{1}{36} \beta \sqrt{g} Q^2  + \hphantom{R}\!\!\!^3R_{ij}Q^{ij} -   \frac{1}{2}  \hphantom{R}\!\!\!^3 R Q\\ & - \frac{1}{4}  Q^{ij}P_{ij} P - \frac{1}{8} Q(P^{ij}P_{ij} -P^2) 
  \end{split}
\ee

The finite energy and momentum configurations correspond to  following asymptotically flat behavior as the boundary condition,
\ben \label{bc} g_{ij} = \delta_{ij} + O(r^{-1}),\quad p^{ij} + O(r^{-2}), \quad Q^{ij} = O(r^{-1}), \quad P_{ij} = O(r^{-1}), \ee 
and that the spatial derivatives of these fields fall off faster by one power of \(r^{-1}\). 

The fields
\be g_{ij} = \delta_{ij} ,\quad Q_{ij}=0,\quad p_{ij}=0 \quad P^{ij}=0 , \ee 
satisfy the constraints and represent a plane in flat space-time. Linearising around this solution, the constraint equations become
\be \partial_i\partial_j Q^{ij} = 0 \quad, \quad \partial_i p^{ij} = 0 ,\ee
while first order changes in \(g_{ij}\) and \(P_{ij}\) remain unconstrained.

The constraint equation is satisfied by \(Q^{ij}=0\) and \(p_{ij}\) to be an arbitrary solution satisfying the boundary condition. Since \(Q^{ij}=0\), energy of the solution given in \eqref{energy} is reduced to
 \be E = \int \{ \frac{1}{2} p^{ij}P_{ij} - \alpha \sqrt{g} C^{0ijk}C^0_{~ijk} \} d^3 x \ee
  \(C^0_{~ijk}\) can be expressed in terms of spatial derivative of \(P_{ij}\). Since \(P_{ij}\) is arbitrary, they choose, \(P_{ij}=\lambda p_{ij}\). Keeping $\lambda$ sufficiently small one can make the second term negligible so that the energy reduces to 
\be E = \int \{ \frac{\lambda}{2} p^{ij}p_{ij} d^3 x .\ee
The sign of the energy will be determined by the sign of the $\lambda$ which could be positive or negative.

They proposed a theorem that  if \(\{g, p, Q, P\}\) satisfy the constraint \( C = 0 \) \eqref{constraint}  and the boundary condition \eqref{bc} with \(\alpha \beta \geq 0 \), then \(E=0\). 

The argument goes as follows. If \(\{g, p, Q, P\}\) satisfy the constraint \( C = 0 \) \eqref{constraint} the boundary condition \eqref{bc} implies that all the terms in the energy in \eqref{energy} fall off faster than \( r^{-2} \), except for the two \( Q^2 \) terms. Thus
\be \begin{split} \label{finalenergy}
 E = - \int  \{ \frac{1}{2} \alpha \sqrt{g} Q^{Tij}Q^T_{~ij} + \frac{1}{36} \beta \sqrt{g} Q^2  + h \} d^3x ,
  \end{split}
\ee
where \( h\) vanishes as fast as \( r^{-3} \).
If  \( E \neq 0\), \( Q^{ij}\) must have a nonzero \(r^{-1}\) contribution and so \( \int Q^{ij} Q_{ij}\) diverges. If \(\alpha\) and \(\beta\) have the same signs the two terms in \eqref{finalenergy} cannot cancel each other. Since the boundary condition requires \( E\) to be finite we encounter a contradiction.

It has also been claimed that this will remain valid in the presence of matter. Since the proof only requires that the constraint equation hold asymptotically, it is required that the energy density of the matter fields fall off faster than \(r^{-2}\).
\bibliographystyle{unsrt}
	\bibliography{reference}

\end{document}